\documentstyle[psfig,rotating]{mn}

\newcommand{\beq}{\begin{equation}}
\newcommand{\eeq}{\end{equation}}
\newcommand{\SSS}{\scriptscriptstyle}
\newcommand{\th}{\thinspace}
\newcommand{\vect}[1]{{\mathbf{ #1}}}
\newcommand{\simgtr}{\mathbin{\lower 2pt\hbox
   {$\rlap{\raise 5pt\hbox{$\char'076$}}\mathchar"7218$}}}
\newcommand{\Msun}{\hbox{$\th {\rm M}_{\odot}$}}
\newcommand{\Rsun}{\hbox{$\th {\rm R}_{\odot}$}}

\title{
Direct N-body Modelling of Stellar Populations: \\ 
Blue Stragglers in M67
}

\author[J. R. Hurley, C. A. Tout, S.J. Aarseth and O. R. Pols]
  {Jarrod R. Hurley$^{1,3}$,
  Christopher A. Tout$^1$, 
  Sverre J. Aarseth$^1$ 
  and Onno R. Pols$^2$\\
  $^1$Institute of Astronomy, Madingley Road, Cambridge CB3 0HA, UK \\
  $^2$ Department of Mathematics, P.O. Box 28M, Monash University, Victoria,
       3800, Australia \\
  $^3$ Current address: Department of Astrophysics, American Museum of 
                        Natural History, Central Park West at 79th Street, 
                        New York, NY 10024 \\ 
  E-mail: {\rm jhurley@amnh.org, cat@ast.cam.ac.uk,
               sverre@ast.cam.ac.uk, onno@mail.maths.monash.edu.au} }

\pagerange{\pageref{firstpage}--\pageref{lastpage}}
\pubyear{2000}

\begin{document}
\label{firstpage}

\maketitle

\begin{abstract}
We present a state-of-the-art $N$-body code which includes a detailed 
treatment of stellar and binary evolution as well as the cluster dynamics. 
This code is ideal for investigating all aspects relating to the evolution 
of star clusters and their stellar populations. 
It is applicable to open and globular clusters of any age. 
We use the $N$-body code to model the blue straggler population of the old 
open cluster M67. 
Preliminary calculations with our binary population synthesis code show 
that binary evolution alone cannot explain the observed numbers or 
properties of the blue stragglers. 
On the other hand, our $N$-body model of M67 generates the required number 
of blue stragglers and provides formation paths for all the various types 
found in M67. 
This demonstrates the effectiveness of the cluster environment in modifying 
the nature of the stars it contains and highlights the importance of 
combining dynamics with stellar evolution. 
We also perform a series of $N = 10\,000$ simulations in order to quantify the  
rate of escape of stars from a cluster subject to the Galactic tidal field. 
\end{abstract}


\section{Introduction}
\label{s:m67int}

The rich environment of a star cluster provides an ideal laboratory for the
study of self-gravitating systems.
It also provides important tests for stellar evolution theory and the
formation of exotic stars and binaries.
The colour-magnitude diagram (CMD) is convenient for displaying 
the range of photometrically observable stellar populations within a 
star cluster. 
However, in the case of dense or dynamically old clusters, the 
appearance of the CMD can be significantly altered by  
dynamical encounters between the cluster stars. 
Therefore it is necessary to combine population synthesis with a 
description of the cluster dynamics. 
This is important for both gravitational and non-gravitational interactions
between stars as well as the dynamical evolution of the cluster as a whole. 
Our approach in this area is to include a consistent treatment of 
stellar and binary evolution in a state-of-the-art $N$-body code so as 
to allow the generation and interaction of the full range of stellar 
populations within a cluster environment (Hurley et al.~2000). 
The stellar evolution algorithm that we use includes variable metallicity 
which enables us to produce realistic cluster models for 
comparison with observed cluster populations of any age. 
As the first group to have this capability we are now embarking on 
a major project to investigate the dynamical evolution of star clusters and 
their populations. 

\begin{figure}
\psfig{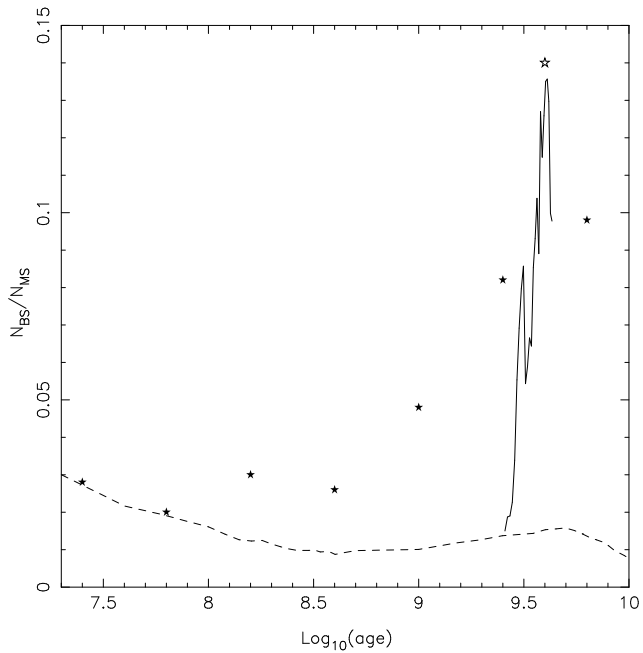}
\caption{
The number of blue stragglers relative to the number of MS stars in the two 
magnitudes below the turn-off as a function of the population age. 
The stars represent the open cluster data of
Ahumada \& Lapasset (1995) with the M67 point an open symbol. 
The dotted line represents our population synthesis with a 50\% 
binary population using the parameters of PS6 (see Section~\ref{s:m67pop}).
The solid line represents our M67 $N$-body simulation 
(see Section~\ref{s:m67nb4}, note that the log-scale does not clearly 
indicate the length of the simulation).  
}
\label{f:fig1bs}
\end{figure}

Blue stragglers are cluster main-sequence (MS) stars that seem to have stayed 
on the main-sequence for a time exceeding that expected from standard 
stellar evolution theory for their mass: 
they lie above and blueward of the turn-off in a cluster CMD. 
Ahumada \& Lapasset (1995) conducted an extensive survey of blue straggler 
candidates in Galactic open clusters. 
Their results are plotted in Figure~\ref{f:fig1bs} as the mean number of 
blue stragglers per open cluster, relative to the number of main-sequence stars 
in the two magnitudes below the turn-off, as a function of the cluster age. 
Also shown in Figure~\ref{f:fig1bs} is the point for the old open cluster 
M67 (NGC$\,2682$) which stands out above the mean for its age, containing 
29 proposed blue stragglers with high probability of cluster membership.  

Milone \& Latham (1992a, hereinafter ML) have undertaken a long-term radial 
velocity observational program to study the blue stragglers of M67. 
Of a total of ten well-observed blue stragglers they find that six are members 
of spectroscopic binaries. 
One of these is a short-period binary; F190 with a period of $4.183\,$d and 
an eccentricity of 0.205 (Milone \& Latham 1992b). 
The others are long-period binaries with periods from 846 to $4913\,$d;  
three have eccentric orbits and two have orbits consistent with being circular 
(Latham \& Milone 1996).
Thus, if the ML sample is taken as representative of the 
overall M67 blue straggler population, then for every ten blue stragglers 
we expect four to be single stars and six to be found in 
binaries with about two of these circular and at least four eccentric.  

This raises the question of how the blue stragglers formed with the most 
obvious scenarios involving binary evolution.
A Case~A mass transfer scenario (Kippenhahn, Weigert \& Hoffmeister 1967) 
involves a main-sequence star filling its 
Roche-lobe and transferring mass to its companion, a less massive 
main-sequence star, followed by coalescence of the two stars as the orbit 
shrinks owing to angular momentum loss. 
The result is a more massive main-sequence star that 
is rejuvenated relative to other stars of the same mass and thus evolves 
to become a blue straggler. 
This is an efficient method of producing single blue stragglers provided that a 
large population of close binaries exists in the cluster.
Case~B mass transfer involves a main-sequence star accreting material from a 
more evolved companion and thus could be a likely explanation for blue 
stragglers in short-period spectroscopic binaries, such as F190.
Blue stragglers in long-period binaries could be produced by Case~C mass 
transfer when the primary is an asymptotic giant branch (AGB) star that has 
lost much of its mass.  
Wind accretion in binaries that initially have fairly large periods could 
also be responsible for such systems.

In all these cases, except perhaps wind accretion, the binary orbits should be 
circularized by tides before and during mass transfer.  
Other scenarios are needed to explain the binaries in eccentric orbits 
and this is where the effects of dynamical interactions in a cluster 
environment become important. 
Physical stellar collisions during binary-binary and binary-single interactions 
can produce blue stragglers in eccentric orbits as well as allowing the 
possibility of an existing blue straggler being exchanged into an eccentric 
binary. 
According to Davies (1996) encounters between binaries and single stars 
become important when the binary fraction in the core exceeds about 5\%
and binary-binary encounters dominate if the fraction is greater
than 30\%.
Additionally, the probability of encounters depends on the density of the
cluster.
It is also possible that perturbations from passing stars may induce an 
eccentricity in a previously circular orbit.
As discussed by Leonard (1996) it is unlikely that any one formation 
mechanism dominates and in the case of the diverse blue straggler population 
of M67 it seems probable that all the above scenarios play a role.

The aim of this paper is to compare $N$-body models with observations of M67 
to investigate the incidence and distribution of blue stragglers (BSs) 
and in so doing to constrain the nature of the primordial binary population. 
In Section~\ref{s:m67obs} we give an overview of the observational data, 
in terms of individual stellar populations and overall cluster parameters. 
We describe the details of our binary population synthesis in 
Section~\ref{s:m67pop} and use it to 
constrain the parameters of the various distributions involved, 
with a view to maximizing the number of blue stragglers produced. 
The $N$-body code is described in Section~\ref{s:nb4cde} and then used in 
Section~\ref{s:m67esc} to quantify the rate of escape of stars from a 
cluster subject to the tidal field of our Galaxy. 
In Section~\ref{s:m67nb4} we present our $N$-body model of M67 paying 
particular attention to the evolution of the blue straggler population. 
Our discussion of the results is then given, followed by conclusions. 

\begin{figure}
\psfig{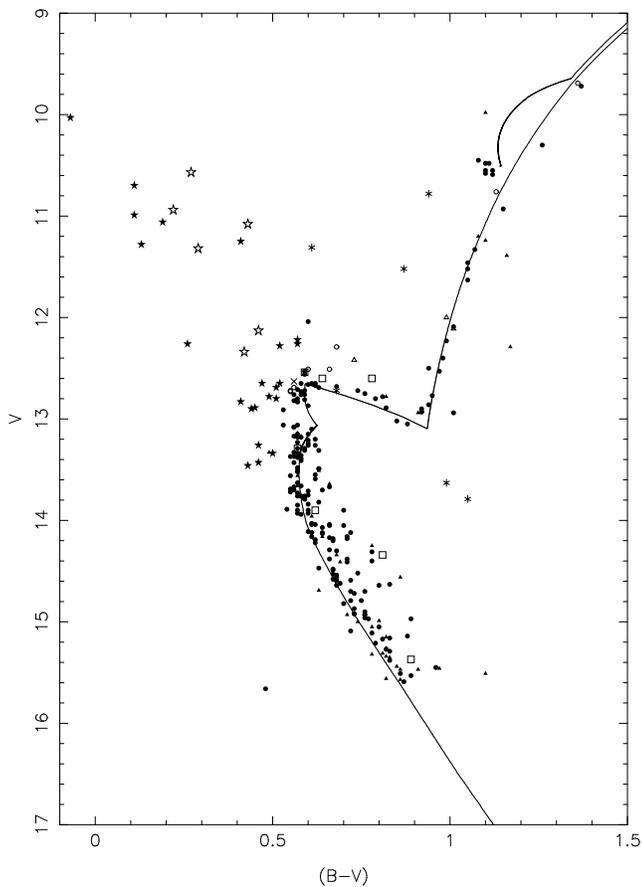}
\caption{
CMD for M67~(NGC$\,2682$) using photometric data taken from the Open Cluster 
Database(OCD: Mermilliod 1996). 
Circles show probable members ($P_{\rm mb} \geq 80\%$) as indicated by 
proper motion studies and triangles show stars of less certain membership. 
Open symbols are spectroscopic binaries. 
Stars identified as blue stragglers in the OCD are plotted with stars. 
Open squares represent the seven RS$\,$CVn candidates identified by 
Belloni, Verbunt \& Mathieu~(1998, BVM). 
Asterisks are the six X-ray sources examined by van den Berg, Verbunt 
\& Mathieu~(1999, MVB), excluding S1082 which has already been plotted as a 
blue straggler.  
The cross represents the triple system observed by Mathieu, Latham 
\& Griffin~(1990, MLG). 
All the special stars that we have highlighted have $P_{\rm mb} \geq 80\%$. 
Also plotted (full line) is an isochrone at $t = 4\,160\,$Myr with $Z = 0.02$, 
$m - M = 9.7$ and $E(B-V) = 0.015$. 
}
\label{f:fg5cmd}
\end{figure}

\section{Observational Data for M67}
\label{s:m67obs}

We use the CCD photometric data of Montgomery, Marschall \& Janes~(1993, 
hereinafter MMJ) taken from the Open Cluster Database 
(OCD: Mermilliod 1996) to construct the M67 CMD shown in Figure~\ref{f:fg5cmd}. 
Proper motion studies by Sanders~(1977) and Girard et al.~(1989) 
distinguish stars with membership probabilities of at least 80\% from 
less certain members.  
The CMD shows a well-defined photometric binary sequence, from which MMJ 
find that at least 38\% of the stars in the cluster are binary systems.  
Of the BSs identified by the OCD, according to the study of 
Ahumada \& Lapasset (1995), eleven are obvious candidates 
from inspection of the CMD. 
These eleven are all in the ML sample which is listed in Table~\ref{t:m67str} 
with each star indicated by its Sanders (1977) number. 
The rest of the BSs are much closer to the MS and its turn-off,  
in a clump with $(B-V) > 0.35$ and $V > 11.8$. 
The two most obvious of these complete the ML sample of thirteen BSs,  
three of which were rotating too rapidly to allow reliable velocity 
determinations.  
Of particular interest among the sample is the super-BS F81 which is a 
single star 
and has a mass of $\simeq 3 \Msun$ (Leonard 1996), more than a factor of 2 
greater than the cluster turn-off mass, $M_{\rm TO} \simeq 1.3 \Msun$. 
The binary S1072 is classified on the OCD as containing a BS but this 
is not compatible with $ubvy$ photometry of the system (e.g. 
van den Berg, Verbunt \& Mathieu 1999, hereinafter MVB) so we do not count  
it as a BS. 
This leaves 28 possible BSs of which two have membership probability less 
than 80\% and almost 2/3 have a membership probability of 98\% or greater. 
Half of the 26 with $P_{\rm mb} \geq 80\%$ are the BSs that have been 
extensively studied by ML, who found that roughly half of their sample were in 
binaries. 

\begin{table*}
\setlength{\tabcolsep}{0.3cm}
\begin{center}
\begin{tabular}{ccccccl} \hline\hline
ID \# & $V$ & $B-V$ & $P$(d) & $e$ & Ref. & comments \\
\hline
S0752 & 11.32 & 0.29 & 1003 & 0.317 & ML & BS binary \\ 
S0968 & 11.28 & 0.13 & & & ML & BS single \\ 
S0975 & 11.08 & 0.43 & 1221 & 0.088 & ML & BS binary \\ 
S0977 & 10.03 & -0.07 & & & ML & BS single (F81) \\ 
S0997 & 12.13 & 0.46 & 4913 & 0.342 & ML & BS binary \\ 
S1066 & 10.99 & 0.11 & & & ML & BS, fast rotator \\ 
S1082 & 11.25 & 0.41 & & & ML & BS single \\ 
S1195 & 12.34 & 0.42 & 1154 & 0.066 & ML & BS binary \\ 
S1263 & 11.06 & 0.19 & & & ML & BS single \\ 
S1267 & 10.57 & 0.27 & 846 & 0.475 & ML & BS binary \\ 
S1280 & 12.26 & 0.26 & & & ML & BS, fast rotator \\ 
S1284 & 10.94 & 0.22 & 4.18 & 0.205 & ML & BS binary (F190) \\ 
S1434 & 10.70 & 0.11 & & & ML & BS, fast rotator \\ 
\hline
S0760 & 13.29 & 0.57 & - & - & BVM & SB, poss. RS$\,$CVn \\ 
S0972 & 15.37 & 0.89 & - & - & BVM & SB, poss. RS$\,$CVn \\ 
S0999 & 12.60 & 0.78 & 10.06 & 0.00 & BVM & RS$\,$CVn \\ 
S1019 & 14.34 & 0.81 & - & - & BVM & SB, poss. RS$\,$CVn \\ 
S1045 & 12.54 & 0.59 & 7.65 & 0.00 & BVM & RS$\,$CVn \\ 
S1070 & 13.90 & 0.62 & 2.66 & 0.00 & BVM & RS$\,$CVn \\ 
S1077 & 12.60 & 0.64 & - & - & BVM & SB, poss. RS$\,$CVn \\ 
\hline
S1040 & 11.52 & 0.87 & 42.83 & 0.027 & MVB & giant + WD \\ 
S1063 & 13.79 & 1.05 & 18.39 & 0.217 & MVB & subsubgiant \\ 
S1072 & 11.31 & 0.61 & 1495 & 0.32 & MVB & BS in OCD \\ 
S1082 & 11.25 & 0.41 & - & - & MVB & BS, poss. companion? \\ 
S1113 & 13.63 & 0.99 & 2.823 & 0.031 & MVB & subsubgiant \\ 
S1237 & 10.78 & 0.94 & 697.8 & 0.105 & MVB & eccentric binary \\ 
S1242 & 12.72 & 0.68 & 31.78 & 0.664 & MVB & eccentric binary \\ 
\hline
S1234 & 12.65 & 0.57 & 4.36 & 0.06 & MLG & triple \\ 
\hline
\end{tabular}
\end{center}
\caption{
Selected M67 stars identified with various observed samples. 
For S1234 the parameters are for the inner orbit. 
}
\label{t:m67str}
\end{table*}

Belloni, Verbunt \& Mathieu (1998, hereinafter BVM) describe the results of 
a ROSAT study of X-ray emission from M67. 
They detect 25 X-ray sources of which three are optically identified with 
circular binaries of orbital periods less than $10\,$d  
suggesting that they are RS Canum Venaticorum (RS$\,$CVn) stars. 
These three are listed in Table~\ref{t:m67str} along with four other X-ray 
sources that are each identified with a spectroscopic binary (SB) that has an 
unknown orbital solution and may possibly be RS$\,$CVn stars. 
Hall (1976) defined RS$\,$CVn systems to have periods between one and 
fourteen days, a hotter component of spectral type F-GV-IV, and strong 
H and K calcium lines seen in emission. 
The H and K emission is generally associated with the cool star which is 
usually a sub-giant. 
Multiply periodic variations in the light curves of these systems have been 
linked to spots on the cool star suggesting enhanced magnetic activity. 
This can be explained by rapid rotation of the cool primary star caused 
by tidal interaction with the orbit which is therefore likely to be 
circular. 
Mass-ratios have been determined for a number of eclipsing RS$\,$CVn systems 
(Popper 1980) and are generally close to one but in several cases are 
greater than one, when $q = M_2/M_1$. 
Here we let $M_1$ denote the primary mass and $M_2$ the secondary mass,  
with the primary defined as the more massive star at formation of the system. 
However, the primary star is not close to filling its Roche-lobe in any of 
these systems so the mass inversion is likely due to a slow mass exchange 
such as wind accretion by the secondary from the primary, or simply rapid 
mass loss from the primary (Tout \& Eggleton 1988). 
X-ray sources in M67 which have not been identified with an optical 
counterpart by BVM are unlikely to be RS$\,$CVn systems because the presence 
of at least one sub-giant or giant would make them highly visible. 
Therefore the seven possible RS$\,$CVn systems is an upper limit to the 
expected number in M67. 

MVB obtained optical spectra for the seven X-ray sources found by BVM 
for which the X-ray emission is unexplained. 
The parameters of these systems are also listed in Table~\ref{t:m67str}. 
One of these sources is the BS S1082 which was determined by 
ML to be single. 
However MVB found a second component in the spectrum of 
this star which they interpret as a hot sub-luminous companion. 
Also among the sample are two so-called {\it subsubgiants} whose nature is 
not yet understood. 

The metallicity given for M67 on the OCD is solar although values 
determined by other authors would indicate that it is slightly 
sub-solar, e.g. [Fe/H] = $-0.04 \pm 0.12$ (Hobbs \& Thorburn 1991) and 
[Fe/H] = $-0.09 \pm 0.07$ (Friel \& Janes 1993). 
The reddening of M67 is reported by various authors to be either  
$E(B-V) = 0.032$ (Nissen, Twarog \& Crawford 1987), $E(B-V) = 0.034 \pm 0.019$ 
(Fan et al.~1996) or $E(B-V) = 0.05$ (MMJ), 
which is rather small considering its distance of about $800\,$pc (OCD) 
from the Sun.  
M67 is an old open cluster with an age of 4 to $5\,$Gyr. 
Carraro et al.~(1994) used $Z \simeq 0.016$, a distance modulus of 
$m - M = 9.5$ and $E(B-V) = 0.02$ to derive an age of $4.8\,$Gyr with 
isochrones based on stellar models that included some convective overshooting.  
Fan et al.~(1996) used a similar metallicity and distance modulus to 
obtain an age of $4.0\,$Gyr from their CCD observations of M67 while the 
OCD gives an age of $5.2\,$Gyr using $m - M = 9.75$. 
The detailed models of Pols et al.~1998, also including convective overshoot, 
used in conjunction with $Z = 0.017$, $m - M = 9.6$ and $E(B-V) = 0.032$ 
give an age of $4.17\,$Gyr. 
From the white dwarf cooling sequence observed in M67 Richer et al.~(1998) 
deduce an age of $4.3\,$Gyr. 
Shown on Figure~\ref{f:fg5cmd} is an isochrone at $t = 4.16\,$Gyr computed 
using the stellar evolution formulae of Hurley, Pols \& Tout~(2000) with 
$Z = 0.02$, $m - M = 9.7$ and $E(B-V) = 0.015$.  
We convert to observed colours with bolometric corrections 
computed by Kurucz~(1992) from synthetic stellar spectra. 
This isochrone gives a good match between the MS hook and the observed 
gap in stars just below the end of the MS. 
It also gives a good fit to the giant branch (GB) and the red clump of 
stars on the zero-age horizontal branch, even with the uncertainties 
in the model atmospheres for cool stars used to compute the colour conversions. 
It should however be noted that an isochrone with $Z = 0.017$, 
$m - M = 9.75$ and $E(B-V) = 0.04$ gives an equally good match for the same 
age and other combinations can be found by varying the age.

\section{Binary Population Synthesis}
\label{s:m67pop}

To investigate the data in Figure~\ref{f:fig1bs} we consider a series of  
binary populations evolved from various initial distributions  
of orbital separations, mass-ratios and eccentricities 
in order to determine the number of blue stragglers produced.  
For each run we evolve two million binaries to an age of $10^{10}\,$yr 
with the binary population synthesis code described by Hurley, Tout 
\& Pols~(2000).  
This code, which represents a thorough reworking of that presented by  
Tout et al.~(1997), includes tidal circularization and synchronization, 
stellar rotation, angular momentum loss mechanisms, common-envelope evolution 
and supernova kicks, in addition to Roche-lobe overflow and mass transfer 
by a stellar wind. 

Initially we vary only the separation distribution, taking the eccentricity 
as uniformly distributed.
We choose the binary mass from the initial mass function (IMF) of 
Kroupa, Tout \& Gilmore (1991, KTG1), by means of the generating function 
\beq\label{e:ktg1mf}
\frac{M_{\rm b}}{\Msun} = 0.33 \left[ \frac{1}{\left( 1 - X \right)^{0.75} 
+ 0.04 \left( 1 - X \right)^{0.25}} - \frac{\left( 1 - X \right)^2}{1.04}
\right] \, , 
\eeq
where X is uniformly distributed between appropriate limits to give 
$0.2 \le M_{\rm b} / \Msun \le 100$. 
Because it has not been corrected for the effect of binaries it is more 
correctly used for total system mass. 
We then choose the component masses according to a uniform distribution of  
mass-ratio constrained by the single star limits of $0.1 M_{\odot}$ and 
$50.0 M_{\odot}$, i.e. 
\beq 
\max \left( 0.1 / \left( M_{\rm b} - 0.1 \right) , 0.02 \left( M_{\rm b} 
- 50 \right) \right) \: \le \: q \: \le \: 1 \, . 
\eeq
We follow Eggleton, Fitchett \& Tout~(1989, hereinafter EFT) by taking the 
distribution of orbital separations expressed as 
\beq\label{e:sepdst} 
\left( \frac{a}{a_m} \right)^{\beta} = \sec\left( kW \right) + 
\tan\left( kW \right) \, , 
\eeq 
where $W \in \left[ -1,1 \right]$, and uniformly distributed and $k$ satisfies 
\beq
\sec k = \frac{1}{2} \left[ {\zeta}^{\beta} + {\zeta}^{- \beta} \right] \, . 
\eeq
This distribution is symmetric in $\log a$ about a peak at $a_m$ and 
ranges from a minimum separation of $\zeta a_m$ to a maximum of $a_m/\zeta$. 
We choose the constants $\zeta$ and $\beta$ to be $10^{-3}$ and $0.33$ 
respectively. 
A choice of $a_m \simeq 30\,$AU corresponds to the Gaussian-like period 
distribution for nearby solar-like stars found by Duquennoy \& Mayor~(1991)  
which has a peak period $P \simeq 180\,$yr. 

The results are shown in Table~\ref{t:m67nbs} where EFT$x$ represents  
separations chosen from eq.~(\ref{e:sepdst}) with a peak, $a_m$, at $x$AU.  
The distribution flat in $\log a$ used in PS4 has the same limits as the EFT10 
distribution of run PS3.
For each run the number of BSs present at $4.2\,$Gyr per $5\,000$  
binaries is shown. 
Our BSs are MS stars that have a mass at least 
2\% greater than the cluster turn-off mass at that time, 
defined as the stellar mass which is currently due to leave the MS. 
Also shown is the number of RS$\,$CVn stars, which may provide  
an additional constraint to the population.  
We define these as circular binaries with $P \leq 20\,$d containing a 
sub-giant or giant primary losing 
mass in a wind, some of which is being accreted by the secondary. 
This ensures that the primary is rotating faster than it would 
as a single star at the same evolutionary stage so that the system is 
magnetically active. 

\begin{table}
\setlength{\tabcolsep}{0.1cm}
\begin{center}
\begin{tabular}{lclcccc} \hline\hline
Run & $Z$ & Separation & $e$ & $q$ & $N_{\rm\SSS BS}$ & $N_{\rm\SSS RS}$ \\
\hline
PS1 & 0.02 & EFT30 & uniform & KTG1, uniform & 1.1 & 1.9 \\
PS2 & 0.02 & EFT17 & uniform & KTG1, uniform & 1.9 & 2.4 \\
PS3 & 0.02 & EFT10 & uniform & KTG1, uniform & 3.2 & 2.7 \\
PS4 & 0.02 & FLAT LOG & uniform & KTG1, uniform & 8.2 & 4.3 \\ \hline
PS5 & 0.02 & EFT10, max 200 & uniform & KTG1, uniform & 3.9 & 3.3 \\
PS6 & 0.02 & EFT10, max 200 & thermal & KTG1, uniform & 4.3 & 3.5 \\
PS7 & 0.02 & EFT10, max 200 & uniform & KTG1, min 0.8 & 3.4 & 3.3 \\
PS8 & 0.02 & EFT10, max 200 & uniform & KTG3 & 2.8 & 2.9 \\
PS9 & 0.01 & EFT10, max 200 & thermal & KTG1, uniform & 5.3 & 3.5 \\
\hline
\end{tabular}
\end{center}
\caption{
Numbers are per $5\,000$ binaries, i.e. $10\,000$ stars, at $4.2\,$Gyr. 
}
\label{t:m67nbs}
\end{table}

We model another five populations using the EFT10 separation distribution 
with an upper limit of $200\,$AU. 
The peak at $10\,$AU in the distribution is still consistent with the 
Duquennoy \& Mayor (1991) observations as is the maximum of $200\,$AU.  
This also agrees with the findings of Mathieu, Latham \& Griffin (1990, 
hereinafter MLG) for 22 spectroscopic binaries in M67 
while the maximum of $200\,$AU is greater than the hard/soft binary 
limit\footnote{Heggie (1975) defined hard binaries to be sufficiently 
close that their binding energy exceeds the mean kinetic energy of the 
cluster stars. For binary component masses similar to the mean stellar mass 
of the cluster this roughly translates to a separation less than 
$a_{\rm hard} = 2 r_{\rm h} / N$ where $r_{\rm h}$ is the 
half-mass radius and $N$ is the number of cluster stars.}  
(Heggie 1975) expected for an open cluster.
Results for runs using EFT10 with varying choices for the eccentricity 
and mass-ratios are also given in Table~\ref{t:m67nbs}.
The IMF used in run PS8 is the single star IMF of Kroupa, Tout \& Gilmore 
(1993, KTG3) and each binary component is chosen independently.
The metallicity for each run is $Z = 0.02$ except for PS9 which examines 
the effect of using a lower value. 

\begin{figure}
\psfig{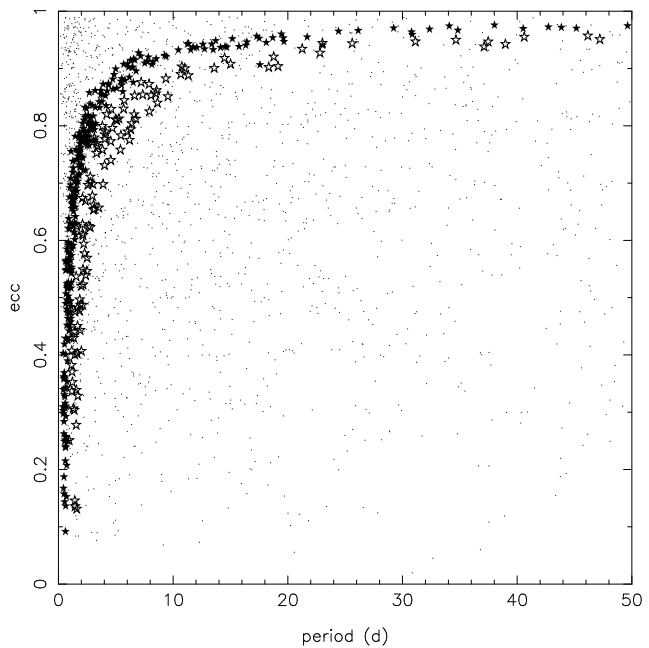}
\caption{
Eccentricity versus Period distribution showing the initial
conditions that evolved to BSs via Case~A mass transfer (solid stars, 
single stragglers) and via Case~B mass transfer (open stars,
stragglers in close circular binaries) for the run PS6. 
Note that some wide Case~C mass transfer binaries were formed at $P > 50\,$d.  
}
\label{f:fig2bs}
\end{figure}

\begin{figure}
\psfig{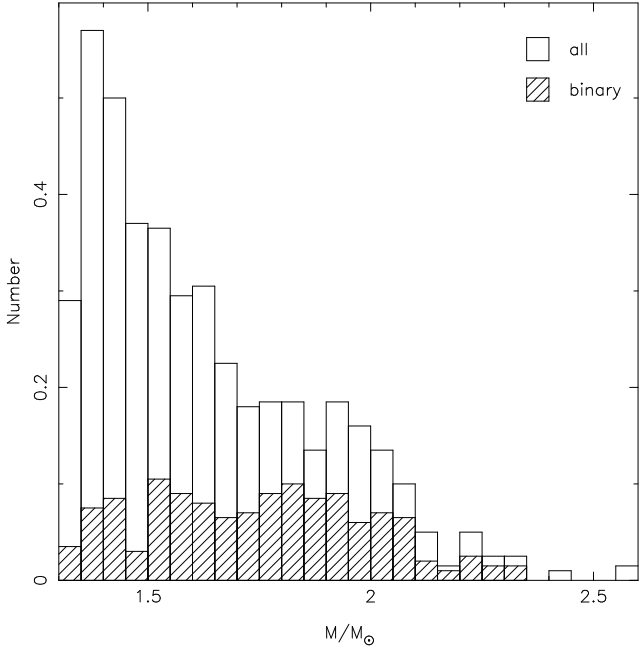}
\caption{
The distribution of stellar mass for blue stragglers at $4.2\,$Gyr in run PS6. 
The distribution is shown for all blue stragglers (hollow) 
and those that are in binaries (hatched). 
The cluster turn-off mass at this time is $M_{\rm TO} = 1.30 \Msun$. 
}
\label{f:figmbs}
\end{figure}

Run PS4 generates the most blue stragglers. 
We show in Section~\ref{s:m67esc} that M67 probably contained about 
$40\,000$ stars initially: roughly $13\,500$ binaries with a 50\% 
fraction. 
If all binaries that produce a BS are retained by the 
cluster then PS4 can explain the number found in M67. 
However, only about 25\% of the BSs are in binaries, all of these 
have circular orbits, and practically none are found in binaries with periods 
greater than a year. 
So dynamical encounters during cluster evolution are required to 
explain BSs found in wide binaries and those found with 
eccentric orbits. 
Although the observations described by Abt~(1983) are consistent with a flat 
distribution of $\log a$, more recent surveys (e.g. Duquennoy \& Mayor 1991) 
favour a peaked distribution such as is used in all runs except PS4. 
The flat distribution has also been ruled out by EFT. 
Of the runs with a peaked distribution, PS6, with a thermal eccentricity 
distribution (Heggie 1975),   
produces the most BSs (excluding run PS9 which has lower metallicity). 
The initial systems in run PS6 that lead to the formation of BSs 
present at $4.2\,$Gyr are shown in Figure~\ref{f:fig2bs}. 
Noticeably evident is the effectiveness of tidal circularization at bringing 
eccentric binaries close enough to make mass transfer possible. 
All instances of Case~A mass transfer lead to coalescence of the two MS 
stars. 
Of the BSs produced by PS6 71.9\% are single, 27.8\% are in close 
binaries resulting from Case~B mass transfer and only 0.3\% are in wide 
binaries produced by Case~C mass transfer or wind accretion. 
Figure~\ref{f:figmbs} shows the mass distribution of BSs present 
at $4.2\,$Gyr. 
Binary evolution alone cannot explain BSs with mass 
greater than twice the cluster turn-off mass, such as F81 in M67. 

The evolution of run PS6 is shown in Figure~\ref{f:fig1bs} for a 50\% 
binary fraction and the assumption that binaries with mass-ratios less than 
0.4 would be observed as MS stars. 
Even though the ratio of BSs to bright MS stars is decreasing initially, 
the number of BSs produced actually increases with time. 
This is because increasingly more stars are evolving off the MS, growing in 
radius and interacting with their companion. 
Also, as time goes on, the masses of the progenitor stars that interact to 
produce BSs decreases so that the BS lifetime increases.  
After some time the number of BSs produced peaks.  
It then starts to fall, mainly as a result of a decreasing number of 
progenitor systems. 
The time of this peak is largely dependent on the 
ratio of the mean binary separation to the size of a star at the MS  
turn-off. 
The latter decreases with time and if stars do not grow to such large  
radii then binary interaction is less likely. 
Additionally, as this ratio increases, stars become more likely to fill their 
Roche-lobes on the GB, if at all, resulting in more cases of common-envelope 
evolution rather than steady mass transfer. 
We find the peak in $N_{\rm\SSS BS}$ for run PS6 occurs at about $5\,$Gyr 
quite unlike Pols \& Marinus~(1994) who found a peak at about $0.1\,$Gyr for 
a separation distribution flat in $\log a$. 
Their mass function was biased towards stars with $M > 2 \Msun$ because they 
were interested in younger clusters. 
At $5\,$Gyr the turn-off mass drops below $1.25 \Msun$ for the first time 
which means that the brighter MS stars begin to have radiative cores 
and therefore do not rejuvenate as much after mass transfer. 
As a result, BS lifetimes decrease in relation to MS lifetimes, contributing to 
the decrease seen in $N_{\rm\SSS BS}/N_{\rm\SSS MS}$ after $5\,$Gyr. 
Stars of lower metallicity have shorter MS lifetimes for $M \la 9 \Msun$ 
so that for a particular age the MS turn-off mass decreases with 
decreasing cluster metallicity. 
Run PS9 has a lower metallicity than PS6 and has a MS turn-off mass of 
$1.24 \Msun$ at $4.2\,$Gyr. 
Consequently the peak in $N_{\rm\SSS BS}$ occurs at about this time which helps 
to explain why more BSs are produced at $4.2\,$Gyr than by PS6. 
Also, stars of the same mass grow to larger radii on the MS for lower 
metallicity so that binary interaction is more likely and 
$N_{\rm\SSS BS}$ in PS9 is always greater than in PS6. 

Figure~\ref{f:fig1bs} shows that, as the population evolves, binary 
evolution alone cannot account for the number of observed BSs  
in open clusters if a realistic separation distribution is used.
Cluster dynamics is therefore not only important for explaining BSs  
found in eccentric and/or wide binaries but is also required to 
increase the number produced. 
We can expect this theoretically through the hardening of primordial binaries, 
which increases the chance of mass transfer, as well as the possibility of 
collisions between MS stars in the cluster core. 
In addition, dynamical evolution together with the effects of a tidal field 
alters the cluster mass function as low-mass stars are stripped preferentially 
from the outer regions (Terlevich 1987).

A problem with all of the population synthesis runs is that the number of 
RS$\,$CVn systems is comparable to the number of BSs, whereas the observations 
suggest that there should only be one for every four BSs. 
This is most likely another signature of the cluster dynamics.  
Giant stars present a much larger cross-section for collisions than MS 
stars and are therefore more likely to be involved in dynamical 
encounters (even though their existence is shorter). 
Moreover, the hardening of close binaries and the disruption of wide binaries 
both act to reduce the number of RS$\,$CVn systems.

\section{The $N$-body Code}
\label{s:nb4cde}

The study of star clusters is currently at a very exciting stage.
Observationally the improved resolution of the Hubble Space Telescope (HST)
is providing a wealth of high-quality information on clusters and their
stellar populations (e.g. Guhathakurta et al. 1998; Piotto et al. 1999).
This includes the dynamically old globular clusters which populate our
Galaxy as well as those of the Magellanic Clouds which
exhibit a wide range of ages.
Coupled to this are the recent advances in computer hardware which have
brought the possibility of direct globular cluster modelling within
reach for the first time (Makino 1999).

In practice the direct integration of an $N$-body system 
presents many technical challenges and has only limited applicability 
to real clusters because the required value of $N$ is too large for a 
simulation to be completed in a reasonable time. 
As a result, most $N$-body simulations performed so far
have involved a varying number of simplified and unrealistic conditions,
such as including only single stars, using only equal-mass stars,
neglecting stellar evolution or assuming no external tidal field 
(e.g. McMillan, Hut \& Makino 1991; Heggie \& Aarseth 1992). 
Despite the fact that, even with simplified conditions, direct $N$-body
simulations are limited to $N$ considerably less than is needed for
globular clusters, much can still be learnt by scaling the results with
particle number, or by modelling small open clusters.
The scaling of time depends essentially on the mechanism to be modelled
(Meylan \& Heggie 1997) but unfortunately the various timescales scale 
differently with $N$ so complications arise when competing processes 
are involved.
Aarseth \& Heggie (1998) discuss these problems further and present a hybrid
time-scaling method which can cope with the transition from early evolution 
related to the crossing time to later relaxation-dominated evolution.

A complication in the use of the results of small-$N$ calculations to make
inferences relating to larger clusters is that many of the structural
properties are $N$-dependent (Goodman 1987).
Also energy considerations which may be dominant for small-$N$, such as the
binding energy of a single hard binary, may not be significant for a larger
system where only the overall energetics is important.
Another problem is that, as $N$ decreases, the results become increasingly
noisy owing to statistical fluctuations.
Casertano \& Hut (1985) have studied how noise affects determination of the
core parameters while Giersz \& Heggie (1994) have suggested that the
situation may be alleviated by averaging the results of many simulations.
The validity of the results of $N$-body calculations has been challenged
on a fundamental level by Miller (1964) who showed that two $N$-body
systems integrated from similar initial conditions diverge exponentially.
Quinlan \& Tremaine (1992) note that this instability occurs on a timescale
comparable with the crossing time, making the results of $N$-body integrations
extremely sensitive to numerical errors and therefore unreliable over
relaxation timescales.
However, they also find that the numerical orbits of $N$-body models are
{\it shadowed} by real systems even though their initial conditions differ.

Ultimately the $N$-body approach remains the method of choice for creating
dynamical models of star clusters because a minimum number of simplifying
assumptions are required and it is relatively easy to implement additional
realistic features.
For a comprehensive description of the various methods for dynamical star 
cluster modelling see Meylan \& Heggie (1997), or Hut et al.~(1992). 

The $N$-body code we use is {\tt NBODY4}, versions of which have been described
in the past by Aarseth (1996, 1999a, 1999b).
This code has been adapted to run on the HARP-3 special-purpose computer
(Makino, Kokubo \& Taiji~1993) and makes use of the many advances made in 
the field since publication of the original direct $N$-body code 
(Aarseth 1963).

\subsection{Integration} 

In $N$-body simulations it is usual to choose scaled units
(Heggie \& Mathieu 1986) such that $G = 1$, the mean mass is $1/N$ and the
virial radius is unity.
This means that, in $N$-body units, the initial energy of the system in virial
equilibrium is $-1/4$ and the crossing time is $2 \sqrt{2}$.
The $N$-body units can be scaled to physical units via the total
stellar mass and an appropriately chosen length-scale factor.
In this work we do not model the initial phase of cluster evolution,
the formation of the cluster.
Instead the initial conditions are chosen subject to suitable observational
constraints and we then integrate a system in virial equilibrium forward
in time.
Basic integration of the equations of motion is performed by the Hermite
scheme (Makino 1991) developed specifically for the HARP.
This scheme employs a fourth-order force polynomial and exploits the fast
evaluation of the force and its first time derivative by the HARP.
It is more accurate than the traditional divided difference formulation
(Ahmad \& Cohen 1973; Aarseth 1985) of the same order and has 
advantages in simplicity and performance.

To exploit the fact that stellar systems involve a wide range of particle
densities, and thus that different particles will have different timescales
for significant changes to their orbital parameters, it is desirable to
introduce individual time-steps.
The individual time-step scheme requires the coordinate and velocity
predictions to be performed for all other particles at each force evaluation.
Fortunately the overheads that this introduces are reduced by utilizing the
HARP hardware for fast predictions to first order.
Significant gains in efficiency are also achieved by adopting quantized 
hierarchical time-steps.
Although first developed by McMillan (1986) to optimize vectorization of
$N$-body codes, the method also aids efficient parallelization of the
force calculations.
The advantage of quantized time-steps is that a block of particles can be
advanced at the same time so that only one prediction call to the HARP is
required for each block.

\subsection{Close Encounters and Regularization} 

When a binary system is within an environment such as a star cluster
where gravitational encounters with other bodies are possible it is not
sufficient to describe the orbit by averaged quantities, as used for 
isolated systems in binary population synthesis (Hurley, Pols \& Tout~2000). 
Instead the orbit must be integrated directly, in a way that
enables the positions of both stars to be known at any time.
This is complicated by the fact that the orbital characteristics are
affected by perturbations from the attractive forces of nearby stars.
If the relative separation of the two binary stars is $\vect{R}$ then
\beq
\ddot{\vect{R}} = - \frac{M_{\rm b}}{R^3} \vect{R} + \vect{P}
\eeq
describes the motion where $\vect{P}$ represents the external tidal
perturbations.
As $\vect{R} \rightarrow 0$ this equation becomes strongly singular, 
leading to increasing errors and small time-steps if integrated by
standard methods.
Therefore alternative methods which regularize the equations of motion for
close encounters, or make use of a hierarchical data structure, 
must be employed.
This includes the case of two strongly interacting particles in a 
hyperbolic encounter.
{\tt NBODY4} makes use of KS regularization (Kustaanheimo \& Stiefel 1965) 
which treats perturbed two-body motion in an accurate and efficient way. 
A recent development has seen the introduction of the Stumpff KS scheme
(Mikkola \& Aarseth 1998) which achieves a high accuracy without extra cost.
This scheme, also incorporating the so-called slow-down principle based on 
adiabatic invariance, is now
used in {\tt NBODY4} and requires about 30 steps per orbit to obtain high
accuracy when relatively weak perturbations are involved.

Since all the KS integrations are carried out on the host machine rather 
than the HARP, and each orbit may require many KS steps, the treatment 
of many perturbed systems is often the most expensive part of the simulation. 
Therefore the extent of the primordial binary population, and its 
distribution of periods, is a major consideration when determining the size 
of a simulation that can be completed in a reasonable time.
For perturbed two-body motion, only contributions from relatively nearby
particles need be considered because the tidal force varies as $1/r^3$.
If no perturbers are selected for a KS pair at apastron then unperturbed KS
motion is assumed and the system can be advanced one or more periods
without stepwise integration.
Because hard (i.e. close) binaries are less likely to be perturbed, a 
significant number of such binaries among the primordial population can 
reduce the load on the host machine.
Thus the choice of period distribution is once again a prime consideration.

To cope with strong interactions between close binaries and single stars the
KS treatment has been extended to three-body regularization
(Aarseth \& Zare 1974).
This basically requires two KS regularizations coupled by suitable
coordinate transformations.
A further generalization to include a fourth body led to the formulation of
chain regularization (Mikkola \& Aarseth 1990, 1993).
The essential feature of this challenging treatment is that dominant
interactions along the chain of particles are modelled as perturbed KS
solutions and all other attractions are included as perturbations.
This is ideal for treating compact subsystems and is currently used in
{\tt NBODY4} for configurations of three to six bodies.

\subsection{Stellar Evolution Treatment}

To provide realistic cluster models the simulation must be able to account
for changes to the radii and masses of stars during the lifetime of a star
cluster.
The evolution of the single and binary stars must be performed in step 
with integration of the dynamics so that interaction between the 
two processes is modelled consistently. 
The earlier version of {\tt NBODY4} included stellar evolution in the form of 
the algorithms presented by Tout et al.~(1997) but this treatment  
is only relevant to stars of Population I composition. 
It is also based on stellar models which have since been superseded by the
models of Pols et al.~(1998) incorporating, among others, improvements to
the equation of state, updated opacity tables, and convective overshooting.
We have now incorporated the updated stellar evolution treatment described by 
Hurley, Pols \& Tout (2000), valid for all metallicities in the range 
$Z = 10^{-4}$ to 0.03, into {\tt NBODY4} in its entirety. 
This means that the metallicity of our cluster models can be varied 
and that a more detailed and accurate treatment of all the single star
evolution phases is used.
Variations in composition can affect the stellar evolution timescales as well
as the appearance of the evolution in a CMD and the ultimate fate of a star. 

All stars carry with them a set of variables that describe their
evolutionary state.
These are the initial mass ($M_0$), current mass ($M_t$), stellar
radius ($R_*$) and stellar type ($k_*$).
Other physical parameters, such as the luminosity and core mass, are not
saved for each star because these are required less frequently
and are calculated when needed.

Because the stars evolve at different rates, depending on their evolutionary
stage and mass, they require different frequencies for the
updating of their variables.
Each star has the associated variables $T_{\rm ev}$ and $T_{{\rm ev}0}$
which represent the next stellar evolution update time,
and the time of the last update, respectively.
Whenever $T \geq T_{\rm ev}$, where $T$ is the simulation time, the star 
is updated by the stellar evolution algorithms.  
The value of $T_{\rm ev}$ is then advanced by a suitable choice of time-step
$\Delta t$ which depends on the evolutionary stage of the star (see
Hurley, Pols \& Tout~2000) and ensures that the physical parameters change
sufficiently smoothly.
After $\Delta t$ is chosen we check whether the stellar radius will change 
by more than 10\% over the interval and reduce $\Delta t$ if this is the 
case. 
In practice the stellar evolution algorithms are not called continuously
because successive calls to the main routine which controls the treatment are
limited by a minimum time interval, which is fairly small.
This means that one call to the routine can involve the updating of many stars
and that an individual star may be advanced several times within one call
if it is in a particularly rapid stage of evolution.

In the simulation all stars have evolved for the same amount of time,
i.e. the age of the cluster, but they may have different relative ages,
for example if a star has been rejuvenated by mass transfer.
Also, when a remnant stage is begun, such as a helium star, white dwarf (WD), 
neutron star (NS) or black hole (BH), the age of the star is reset to zero 
for the new type.
Therefore the additional variable {\sc epoch} is introduced for each
star so that its age at any time is given by the difference between the 
current cluster age and its {\sc epoch}. 

We include mass loss by stellar winds according to the prescription
given in Hurley, Pols \& Tout (2000).
The time-step is limited so that a maximum of 2\% of the stellar mass can be
lost.
Any mass lost which is not accreted by a close companion is assumed to leave 
the cluster instantaneously, with the appropriate corrections made to account
for the change in potential energy, maintaining energy conservation
for the system.
The force on nearby stars must also be modified as a result of the mass
change or, if a significant amount of mass is lost, a complete
re-initialization of the force polynomials on the HARP may be required.
Also, if a NS or BH is formed, a velocity kick taken from a Maxwellian
distribution with dispersion $\sigma = 190\, {\rm km} \, {\rm s}^{-1}$ 
(see Hurley, Tout \& Pols~2000) is given to the supernova remnant
which is usually enough to eject it from the cluster.

\subsection{Binary Evolution Treatment}

For the treatment of evolution within a binary system we include the 
features described in Hurley, Tout \& Pols~(2000) over and above those of 
the old algorithms (Tout et al.~1997).
However, we use the treatment of tidal circularization developed by
Mardling \& Aarseth (2000, hereinafter MA) 
which includes features that cope with the added complication of external
perturbations to the orbital parameters, and has been implemented to work
closely with the KS scheme.

Each binary has an associated composite particle, the centre-of-mass (CM) 
particle, which has its own set of variables, $M_t$, $k_*$, 
$T_{\rm ev}$ etc., as well as additional variables such as the binding 
energy per unit mass of the system. 
A typical binary will make a journey through a series of CM evolution
stages, beginning with $k_{*,{\rm{\SSS CM}}} = 0$ when 
it is first created: either primordially or during the evolution.
A close binary orbit circularizes tidally as  
energy is dissipated but angular momentum is conserved. 
Tidal circularization ($k_{*,{\rm{\SSS CM}}} = -2$) is activated when the
circularization timescale of the binary is less than $2 \times 10^9\,$yr.
Note that the intrinsic stellar spin is not incorporated in the 
circularization treatment so that the angular momentum of an 
unperturbed binary orbit remains constant during the circularization process. 
This could be rectified in line with the tidal model of 
Hurley, Tout \& Pols~(2000).

For a binary of high eccentricity it is possible for the energy exchange
between the orbit and the tides at periastron to become chaotic 
($k_{*,{\rm{\SSS CM}}} = -1$).
This means that oscillations occur which gradually damp the system until
sufficient energy is dissipated for the {\it chaos boundary} to be crossed,
when the orbit settles on some eccentricity whence angular momentum is
conserved, and the orbit begins to circularize (MA). 
The implementation is based on a non-linear dissipation timescale
and assumes conservation of total angular momentum for the system 
(i.e. the stars spin up).
Chaotic behaviour is most likely to occur in tidal-capture binaries and
MA note that it is still not clear how stars in chaotic orbits 
respond to the huge tidal energies involved.
Tout \& Kembhavi (1993) and Podsiadlowski (1996) have shown that the response 
of the tidally heated
primary depends on the region of the star in which the energy is deposited
and also the timescale on which this energy is thermalized within the star.

Once the circularization is complete ($k_{*,{\rm{\SSS CM}}} = 10$) 
the binary is tested regularly for Roche-lobe overflow (RLOF). 
Prior to circularization the unlikely possibility of RLOF is ignored.
In most cases this technicality does not lead to any physical irregularities
because tidal interaction generally acts to remove any eccentricity on a 
timescale shorter than the evolution timescale of the binary.
However problems may arise if a system forms in a close eccentric orbit.  
Even then there is evidence that some circularization occurs during the 
formation process (Mathieu 1994).
While the orbit is circularizing the stars are
also evolving and growing in radius, so contact is expected in some
systems as the orbit approaches zero eccentricity.
However the timescale for a collision is actually prolonged because the 
periastron distance grows as $e$ shrinks,
owing to conservation of angular momentum.
Thus, as the stars are expanding so is the minimum separation between them.
Also most binaries do not begin to circularize, or to come close to a RLOF
state, until one of the stars is on the GB or AGB, at which point it is losing
mass in a stellar wind that causes an additional increase in the separation.
Nevertheless, stages of rapid radius growth may result in premature 
coalescence. 

It is also possible for the eccentricity of a binary to
increase as a result of external perturbations acting on the orbit.
If this induces an eccentricity in an orbit which was previously
circular the binary is reset to standard type ($k_{*,{\rm{\SSS CM}}} = 0$). 
This is also done if an eccentric binary survives one of the component
stars exploding as a supernova to leave a NS or BH remnant.

While a binary is in a detached state the stars are updated 
in the same way as single stars.
The only difference is that both stars must be treated at the same
time so that the possibility of wind accretion can be modelled.
If the CM type is $k_{*,{\rm{\SSS CM}}} = 10$ or greater, each time the
binary stars are updated the time until the primary star fills its 
Roche-lobe is estimated. 
This is used to set $T_{\rm ev, {\SSS CM}}$.
If at any stage $T \geq T_{\rm ev, {\SSS CM}}$ then RLOF could have begun
and the binary is subject to the {\sc roche} procedure 
($k_{*,{\rm{\SSS CM}}} = 11$)
where it is evolved forward according to the treatment described in
Hurley, Tout \& Pols~(2000). 
This means that the physical time of the binary evolution can move slightly 
ahead of the cluster integration time, introducing the need for coasting 
periods in which the binary is put on hold until the dynamical 
integration time catches up.
The amount of time that the binary moves ahead during {\sc roche} is determined
by physical conditions of the components.
Any stellar wind mass loss during RLOF is also dealt with inside {\sc roche}
so it is important that $T_{\rm ev}$ for each of the component stars is
greater than the time reached by the binary when it exits the active
RLOF stage, i.e. $T_{\rm ev, {\SSS CM}} < T_{\rm ev,1} , T_{\rm ev,2}$.
In this way the component stars have their evolution treated as part
of the RLOF process and not as individuals.
A sustained phase of RLOF may involve a series of active {\sc roche} calls
each followed by a coasting period.
It is also possible that a binary may experience more than one phase of
RLOF during its evolution.
This is easily accommodated by the treatment.
During mass transfer the separation of the binary stars changes and the
KS variables are updated frequently.
As a fundamental variable for two-body regularization the binding energy
per unit mass must be determined accurately at all times.
In fact, all the KS variables must be corrected for mass loss. 
A mass-loss correction is also made to the total energy of the cluster and the
force polynomials of nearby stars are re-initialized.

We treat common-envelope evolution and collisions that arise as a result of 
contact systems as part of the {\sc roche} process.
In general, direct stellar collisions during the cluster evolution are
very rare because the interacting stars most likely form a tidal capture
binary, which may be quickly followed by coalescence via a common-envelope
or contact phase.
This is due to the relatively low velocity dispersion, $\sigma \simeq 10 \,
{\rm km} \, {\rm s}^{-1}$, of cluster stars.
In galactic nuclei where typically $\sigma \simeq 100 \, {\rm km} \,
{\rm s}^{-1}$ direct collisions are expected.
We use a periastron criterion, derived for main-sequence stars
(Kochanek 1992), 
\beq
a \left( 1 - e \right) \: < \: 1.7 \left( \frac{M_1 + M_2}{2 M_1}
\right)^{1/3} \, R_1 \, ,
\eeq
where $R_1$ is the primary radius, to determine direct collisions.

Because common-envelope events and collisions frequently lead to 
coalescence, either one or two particles can become redundant. 
Similarly some single star supernovae may not leave a remnant.
In such cases it is simplest to create a massless component and place it well 
outside the cluster so that it escapes soon. 
If the component of a binary is removed then the coalescence product is given 
the CM coordinates and velocity of the original binary.
Energy corrections associated with mass loss are performed and a new force
polynomial initialized.

As already mentioned, the fraction of the stars in primordial binaries is 
of great importance. 
It is generally accepted that some degree of primordial population is present 
because the rate of binary formation during the evolution would not be
enough to halt core-collapse appreciably (Hut et al.~1992). 
Binary stars in clusters can also be identified through their position in the
CMD of the cluster as the combination of light from
the two stars displaces the binary from the position of a single star
having the same mass as either of the components (Hurley \& Tout 1998). 
This is most noticeable on the MS where the existence of a distinct binary
sequence has been exploited by the resolution of the Hubble Space Telescope
(HST) to reveal globular cluster binary fractions of 10 to 30\%
(e.g. Richer et al. 1997; Elson et al. 1998).
For open clusters results are often uncertain because the number
of stars is smaller and membership can be difficult to determine.
Even so, a significant number of pre-main-sequence binaries and multiple
systems have been found in young open clusters and associations
(e.g. Simon et al. 1995; Brandner et al. 1996). 

\subsection{Hierarchical Systems}

During the evolution of a star cluster stable
multiple systems can form as a result of dynamical interactions.
The formation of hierarchical triples, in which one component of a binary
is itself a binary, has been shown to occur in various scattering experiments
(e.g. Mikkola 1983; Bacon, Sigurdsson \& Davies 1996) mainly as a result of
strong binary-binary encounters.
The inner binary of such a system may be relatively hard so that
integration of the strongly perturbed KS solution can prove to be
extremely time-consuming.
This is especially true if the hierarchy is stable for a long period of time.
In that case the inner binary only experiences
short-term fluctuations in its orbital parameters, and so it is
acceptable to perform direct integration only of the outer orbit while the
system remains stable.

A semi-analytical stability criterion based on the analogy with the
chaos boundary in tidal evolution has been presented by MA and
implemented in {\tt NBODY4}.
They employ a critical periastron distance for the outer orbit in terms
of the inner semi-major axis, $a_{\rm in}$, given by
\beq
R_{\rm p, crit}^{\rm out} = C \left[ \left( 1 + q_{\rm out} \right)
\frac{\left( 1 + e_{\rm out} \right)}{\left( 1 - e_{\rm out} \right)^{1/2}}
\right]^{2/5} \, a_{\rm in} \, ,
\eeq
where $e_{\rm out}$ is the eccentricity of the outer orbit and $C \simeq 2.8$
is determined empirically.
The mass-ratio of the outer orbit is $q_{\rm out} = M_3 / \left( M_1 + M_2
\right)$, where $M_1$ and $M_2$ are the component masses of the inner orbit.
If the outer periastron separation is greater than $R_{\rm p, crit}^{\rm out}$
we consider the hierarchy to be stable and temporarily merge the stars  
into one KS system consisting of the CM particle of the inner binary and the
third body, $M_3$.
Because the period of the outer orbit is considerably longer than that of the
inner orbit this procedure greatly reduces the computational cost.
Quadruple and higher-order systems are similarly dealt with.
Procedures to model the cyclic oscillations of the inner eccentricity,
the Kozai effect (Kozai 1962), and any tidal circularization
that is induced, are implemented according to MA.

The possibility of an exchange interaction, in which one of the inner binary
components is displaced by an incoming third star, is checked by the 
criterion of Zare (1977)\footnote{Because of the degeneracy of angular 
momentum, this criterion is only used for small inclinations}.
As noted by Aarseth (1999a), the stability boundary lies above the
exchange boundary when all the masses involved are comparable and the two
only begin to overlap when $q_{\rm out} \simeq 5$.
If an exchange does occur then the expelled star invariably leaves the
three-body system altogether.

To ease the book-keeping required when particles are regularized or
hierarchies are formed, the simulation particles are kept in an ordered data
list throughout the evolution.
Consider a model composed of $N_{\rm s}$ single stars and $N_{\rm b}$
regularized binaries, i.e. $N = N_{\rm s} + 2 N_{\rm b}$ stars in total.
The first $2 N_{\rm b}$ entries in the data list are the binary stars, with
each binary pair grouped together, followed by the $N_{\rm s}$ single star
entries, and finally the $N_{\rm b}$ CM particles.
Take the case of a basic KS binary which is combined with a single star to
form a stable hierarchy.
The single star is moved to the position formerly occupied by the
second component of the binary and the CM particle of the old binary
is moved to the first component position.
The values, such as binding energy, of the old binary CM are moved to
the position formerly occupied by the single star which now has zero mass and
is a {\it ghost} particle, i.e. it does not contribute to force calculations.
Component masses of the inner binary are saved in a merger table and
the CM position for the binary now holds variables relevant to the outer
orbit.
If the hierarchy is broken up then all the variables are easily 
re-assigned and the inner binary restored.
Possible reasons for termination of a hierarchy include violation of the
stability criterion or the onset of RLOF in the inner binary.
If the hierarchy involves the merger of two binaries then both CM particles
would be the components of the new binary.

\section{Escape from a Tidally-Limited Cluster}
\label{s:m67esc}

Fan et al. (1996) present CCD spectrophotometry of $6\,558$ stars in the field 
of M67.
They estimate that the MS is complete down to $0.5 M_{\odot}$ 
and that the cluster has a total observed mass of approximately 
$1\,000 M_{\odot}$ in stars with masses greater than $0.5 M_{\odot}$. 
The observations reveal a mean (projected) half-mass radius of $2.5\,$pc for  
MS stars compared with $1.6\,$pc for the (more centrally condensed) 
BS population.  
The tidal radius is at $10\,$pc and their data is consistent with a 
50\% binary fraction.

With an initial population comprising single star masses chosen from the 
KTG3 IMF, binary masses from the KTG1 IMF and a 50\% binary fraction,  
about $7\,000$ stars ($3\,500 M_{\odot}$) are required to give a current  
cluster mass of $2\,500 M_{\odot}$ (corresponding to $1\,000 M_{\odot}$ 
observed above $0.5 M_{\odot}$).
The mass loss is due solely to stellar evolution and assumes no dynamical  
effects on the population.
However, in a cluster environment stars undergo gravitational 
interactions so that from time to time an encounter gives enough energy 
to a star that it can escape from the system.
The timescale over which the stars evaporate in this way is related to the  
relaxation timescale, $t_{\rm r}$, of the cluster and is shortened by the 
presence of an external tidal field as well as the inclusion of a full 
spectrum of stellar masses.

An isolated spherical cluster with potential $\Phi$ has an escape speed 
$v_{\rm e}$ at radius $r$ given by 
\beq
v_{\rm e}^2 = - 2 \Phi \left( r \right) \, . 
\eeq
The mean-square escape speed in a system with uniform density is 
\beq 
\langle v_{\rm e}^2 \rangle = 4 \langle v^2 \rangle \, , 
\eeq
so the RMS escape speed is twice the particle RMS speed. 
For a Maxwellian velocity distribution the fraction of particles that have 
speeds exceeding twice the RMS speed is $\gamma = 7.4 \times 10^{-3}$ 
(see Binney \& Tremaine 1987, p.~490) and  
the evaporation process can be approximated by removing 
a fraction $\gamma$ of stars each relaxation time, i.e. 
\beq
\frac{dN}{dt} = - \gamma \, \frac{N}{t_{\rm r}} \, . 
\eeq
Evaporation then sets an upper limit to the lifetime of the bound stellar 
system of about $10^2 t_{\rm r}$. 
Note that numerical solution of the Fokker-Planck equation gives 
$\gamma = 8.5 \times 10^{-3}$ (Spitzer 1987, p.~54). 
The total energy of a cluster with total mass $M$ and radius $R$ is, 
according to the virial theorem, 
\beq\label{e:clsteg}  
E = - k \frac{G M^2}{R} \, , 
\eeq
where $k$ is a dimensionless constant of order unity. 
If it is assumed that the cluster evolution is self-similar, so that its shape 
remains fixed, $k$ is independent of time. 
Since evaporation is mostly driven by weak encounters the stars escape 
with very little energy so that $E$ remains essentially fixed. 
Therefore 
\beq
\frac{r}{r_0} = \left( \frac{M}{M_0} \right)^2 
\eeq
and the cluster contracts as it loses mass. 

The evolution of a real star cluster is somewhat different to that of an 
isolated uniform system because the cluster is subject to the tidal force  
of the galaxy in which it resides. 
Consider a cluster in a circular orbit around a galaxy at a distance  
$R_{\rm\SSS G}$ from the galactic centre. 
To estimate the strength of the tidal field exerted on the cluster we assume 
that $M_{\rm\SSS G}$, the galactic mass enclosed by the orbit, is 
distributed throughout the inner spherical volume. 
By choosing the origin of a rotating reference frame to be the cluster 
centre-of-mass, the $x$-axis directed away from the galactic centre, 
the $y$-axis in the direction of rotation, and linearizing the tidal 
field, the equations of motion are 
\begin{eqnarray} 
\ddot{x} & = & F_x + 2 {\omega}_{\rm\SSS G} \dot{y} + 
3 {\omega}_{\rm\SSS G}^2 x \nonumber \\ 
\ddot{y} & = & F_y - 2 {\omega}_{\rm\SSS G} \dot{x} \\ 
\ddot{z} & = & F_z - {\omega}_{\rm\SSS G}^2 z 
\end{eqnarray} 
(Giersz \& Heggie 1997), where 
\beq
{\omega}_{\rm\SSS G} = \sqrt{\frac{G M_{\rm\SSS G}}{R_{\rm\SSS G}^3}} 
\eeq
is the angular velocity. 
The tidal radius $r_{\rm t}$ of the cluster is defined by the 
saddle point on the $x$-axis of the effective cluster potential, 
analogous to the definition of the Roche-lobe radius in a binary system 
(except that the cluster is not co-rotating with its orbit). 

The Galactic tidal field can conveniently be described in terms of Oort's 
constants, 
\begin{eqnarray*} 
A & = & 14.5 \pm 1.5 \: {\rm km} \, {\rm s}^{-1} \, {\rm kpc}^{-1} \\ 
B & = & -12 \pm 3 \: {\rm km} \, {\rm s}^{-1} \, {\rm kpc}^{-1} 
\end{eqnarray*} 
(Binney \& Tremaine 1987, p.~14), so that 
\beq
A - B \: = \: \left( \frac{v_{\rm\SSS G}}{R_{\rm\SSS G}} \right)_{R_0} \: = \: 
26.5 \pm 4 \: {\rm km} \, {\rm s}^{-1} \, {\rm kpc}^{-1} \, . 
\eeq
This is consistent with an orbital velocity of $v_{\rm\SSS G} = 220 
\: {\rm km} \, {\rm s}^{-1}$ (Chernoff \& Weinberg 1990) at $R_0 = 8.5\,$kpc. 
In this formulation the tidal radius is 
\beq\label{e:r2tide} 
r_{\rm t} = \left( \frac{G M}{4 A \left( A - B \right)} \right)^{1/3} \, . 
\eeq 

The cluster escape rate can be expressed by 
\beq\label{e:escrte}  
\frac{dM}{dt} = - \frac{k_{\rm e}}{{\log}_{10} \Lambda} \frac{M}{t_{\rm rh}}
\eeq 
where $M$ is the cluster mass at time $t$ for $N$ stars, 
$\Lambda = 0.4 N$ is used in the Coulomb logarithm and 
\beq\label{e:trh5ch} 
t_{\rm rh} = 0.894 \, \frac{N}{{\log}_{10} \Lambda} \, \frac{r_{\rm h}^{3/2}}
{M^{1/2}} \, \, {\rm Myr}  
\eeq 
is the half-mass relaxation time if $r_{\rm h}$ is in pc and $M$ in $M_{\odot}$.
The constant $k_{\rm e}$ quantifies the rate of mass lost in stars stripped 
from the cluster, where $\gamma = k_{\rm e}/{\log}_{10} \Lambda$ because 
$M \propto N$. 
Numerical solution of the Fokker-Planck equation for a tidally truncated 
cluster gives $\gamma = 4.5 \times 10^{-2}$ (Spitzer 1987, p.~59). 

\begin{figure}
\psfig{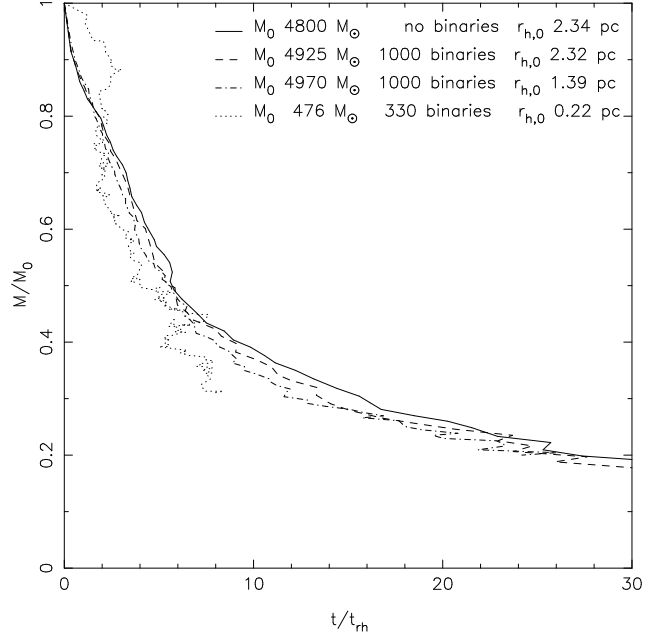}
\caption{
Evolution of the total mass of bound stars for $N$-body models
started with $10\,000$ stars and a standard tidal field with the time
scaled by the current half-mass relaxation time of the cluster. 
Also shown is the evolution for
a model started with only $1\,000$ stars but with a 50\% binary fraction.
}
\label{f:fig3bs}
\end{figure}

\begin{figure}
\psfig{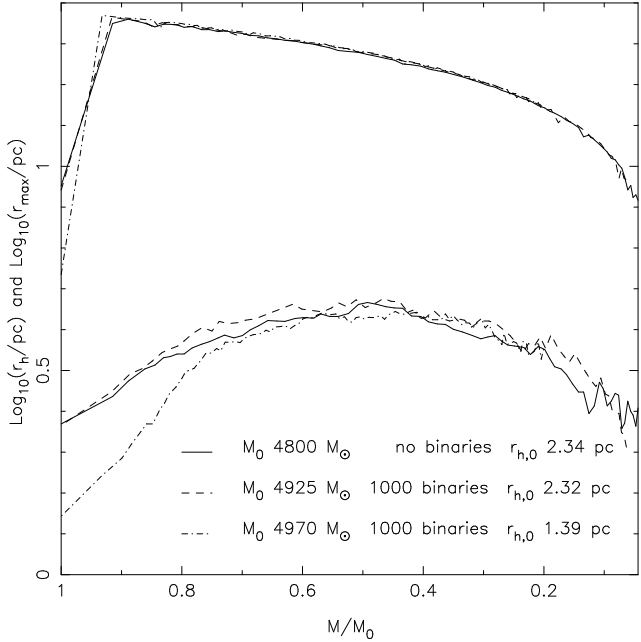}
\caption{
Evolution of the maximum and half-mass radii for the same 
$N = 10\,000$ models as in Figure~\ref{f:fig3bs}. 
Note that the maximum radius is given by the radius of the star closest 
to the tidal radius. 
It quickly approaches the tidal radius following the initial expansion of 
the cluster.  
Since stars are not removed from the cluster until $r > 2 r_{\rm t}$ there 
is a small population of cluster members with 
$r_{\rm max} < r < 2 r_{\rm t}$. 
On average the mass in cluster stars outside the tidal radius is less than 
2\% of the total cluster mass. 
}
\label{f:fig4bs}
\end{figure}

We have investigated cluster escape rates for simulations with a full mass 
spectrum, mass loss from stellar evolution and a standard Galactic tidal field 
($v_{\rm\SSS G} = 220 \, {\rm km} \, {\rm s}^{-1}$ and 
$R_{\rm\SSS G} = 8.5\,$kpc), 
in a series of $N$-body models evolved with {\tt NBODY4}. 
In these simulations we assume the stars are stripped from the  
cluster when $r > 2 r_{\rm t}$. 
This underestimates slightly the escape rate compared to letting 
them escape when they are outside the tidal radius (Giersz \& Heggie~1997). 
We performed a total of six simulations starting with $N = 10\,000$. 
In each model the initial positions and velocities of the stars are 
assigned according to a Plummer model (Aarseth, H\'{e}non \& Wielen 1974) 
in virial equilibrium with the tidal radius defined by eq.~(\ref{e:r2tide}). 
The single star masses are chosen from the KTG3 IMF and the binary masses 
from the KTG1 IMF, with minimum and maximum single star limits of 
$0.1$ and $50 \Msun$. 
We set the distribution of mass-ratios for the binaries to be uniform between 
0.1 and 1 and the metallicity at $Z = 0.02$. 
We choose the binary separations from the same distribution as 
population synthesis run PS1. 
Three of the simulations had no primordial binaries and a length scale 
chosen so that $r_{\rm h,0} \simeq 0.1 r_{\rm t,0}$. 
The other three had a primordial binary fraction of 
$f_{\rm b} = 0.11$\footnote{We define the binary fraction as 
$f_{\rm b} = N_{\rm b}/(N_{\rm b} + N_{\rm s})$ for $N_{\rm b}$ binaries 
and $N_{\rm s }$ single stars. During this work we may also refer to 
a binary fraction as a percentage, i.e. $f_{\rm b} = 50\%$, and we 
note that this should more correctly be called a binary frequency.}: two  
of these with the same initial half-mass radius as the single star models 
and one, started with $r_{\rm h,0} \simeq 0.06 r_{\rm t}$, more condensed. 
In each simulation the initial cluster mass, $M_0$, is about $5\,000 \Msun$ 
which gives a tidal radius of about $22\,$pc. 

The evolution of the cluster mass as a function of the number of half-mass 
relaxation times is shown in Figure~\ref{f:fig3bs}.
The inclusion of binaries has little effect, nor does 
varying the initial concentration of the cluster, as long as the time is 
scaled by the current half-mass relaxation time.
In fact the initial size has little effect because, owing to mass loss from 
massive stars and core contraction, the cluster quickly expands to fill 
its tidal radius which depends only on mass. 
This is demonstrated in Figure~\ref{f:fig4bs} which shows the variation of 
the half-mass radius with cluster mass as well as the position of the 
star closest to the tidal radius at the time. 
As each cluster evolves, the half-mass radius, which initially expands freely, 
starts to feel the effect of the contracting tidal radius so that the rate 
of expansion slows. 
Eventually the half-mass radius begins to contract and the cluster 
evolution becomes remarkably self-similar. 
For each simulation the turn-over occurs at $M/M_0 \simeq 0.5$ when 
$t/t_{\rm rh} \simeq 6$. 
During the subsequent self-similar evolution phase the ratio of the tidal and 
half-mass radii is $r_{\rm t}/r_{\rm h} \simeq 4$. 

The mean core-collapse (CC) time for all the models is $t_{\rm\SSS CC} 
\simeq 4 t_{\rm rh}$. 
In terms of the initial half-mass relaxation time this is 
$t_{\rm\SSS CC} \simeq 7 t_{\rm rh,0}$ because $t_{\rm rh,{\SSS CC}} \simeq 
2.2 t_{\rm rh,0}$ owing to the initial expansion of $r_{\rm h}$. 
At the time of core-collapse an average of five hard binaries have formed 
in the core of the single star models. 
The model with the smaller $r_{\rm h,0}$ reached the $r_{\rm h}$ turn-over 
point in about 85\% of the time it took the lower density models but this 
single result is not significant.  
 
From the $N = 10\,000$ models we find that $k_{\rm e} \simeq 0.3$ 
fits the data adequately for almost the entire cluster evolution. 
This result is consistent with a series of six $N = 1\,000$ simulations 
that we began with $f_{\rm b} = 0.5$ and $r_{\rm h,0} \simeq 0.03 r_{\rm t}$. 
For these smaller models the half-mass radius starts to contract at 
$M/M_0 \simeq 0.5$. 
The mass-loss rate is also compatible with the results of Giersz 
\& Heggie~(1997) from a series of $N$-body models with $N = 500$, when we  
take into account the fact that Plummer models have a lower escape rate 
than King models. 
This is due to Plummer models having a slightly weaker tidal field,  
and therefore evolving more slowly in the post-core-collapse phase of 
evolution, than their King model counterparts. 
Giersz \& Heggie~(1997) use a minimum mass of $0.4 \Msun$, compared to the 
value of $0.1 \Msun$ in the models presented here, so they have a 
higher proportion of massive stars. 
Their shorter evolution timescales cause the simulations to evolve faster 
which increases the escape rate and partly explains why their value of 
$k_{\rm e} \simeq 0.6$ (converted to the units used here assuming 
$\Lambda = 0.4N$ for both sets of models) is larger. 
Giersz \& Heggie~(1997) found that $\Lambda = 0.015N$ is required 
to make the results of their models agree with the Fokker-Planck models of 
Chernoff \& Weinberg~(1990), in which case the conversion of their escape 
rate gives an even larger value of $k_{\rm e} \simeq 1.5$.  

So how robust is our $k_{\rm e}$ against changes to the initial conditions 
assumed for the $N$-body models? 
We have already discussed that it is sensitive to the model from 
which the density and velocity profiles are taken: Plummer, King or 
otherwise. 
The mass function used also has an effect even though a change 
in IMF slope will be offset to some degree by the inverse 
dependence of the relaxation timescale on average stellar mass. 
Our escape rate has been derived by considering a number of model sizes 
and binary fractions but must be tested over a greater range of 
parameter space before it can be universally accepted. 
Factors such as the proportion of hard binaries must also have an, 
as yet undetermined, effect on the result. 
Caution should be exercised when applying $k_{\rm e}$ to situations 
where the initial conditions differ substantially with those from which 
it was derived. 

If the evolution of the cluster is self-similar then 
\beq\label{e:radevn} 
\left( \frac{r_{\rm h}}{r_{\rm h,0}} \right) = \left( \frac{M}{M_0} 
\right)^{2 - \zeta} \, , 
\eeq
which stems from writing the rate of change of the total cluster energy as 
\beq 
\frac{dE}{dt} = \frac{\zeta E}{M} \frac{dM}{dt} \, . 
\eeq
If $\zeta > 2$, $r_{\rm h}$ increases as $M$ decreases.  
This can occur during the initial violent relaxation phase when stellar wind 
mass-loss causes an overall expansion, or during the post-core-collapse 
expansion of the inner regions. 
If $\zeta = 5/3$ then the evolution of the half-mass radius follows the 
decrease in tidal radius as the cluster evolves. 
Using eq.~(\ref{e:radevn}) to integrate eq.~(\ref{e:escrte}) gives 
\beq\label{e:mtmngt} 
M \left( t \right) = M_0 \left[ 1 - \frac{7 - 3 \zeta}{2} 
\frac{k_{\rm e}}{{\log}_{10} {\Lambda}_0} \frac{t}{t_{\rm rh,0}} 
\right]^{\frac{2}{7 - 3 \zeta}} \, . 
\eeq
We can use this with $k_{\rm e} = 0.3$ to estimate the initial mass and 
half-mass radius required to give the current values of $M \simeq 2\,500 \Msun$ 
and $r_{\rm h} \simeq 2.5\,$pc at $t \simeq 4\,200\,$Myr for M67. 
However, as Figure~\ref{f:fig4bs} shows, a complication arises because $\zeta$ 
as it is defined in eq.~(\ref{e:radevn}) is not constant throughout the 
cluster lifetime.  
Furthermore, the evolution of a tidally-limited cluster is not 
self-similar initially.  

The initial cluster values can be roughly constrained by the 
following method. 
For $M/M_0 \leq 0.5$ the $N = 10\,000$ 
models show that $\zeta = 5/3$ and $r_{\rm t} = 4 r_{\rm h}$. 
So we choose a value for $M_0$ and use eq.~(\ref{e:r2tide}) to find 
$r_{\rm t,1/2}$, i.e. tidal radius when $M = 0.5 M_0$. 
This defines $r_{\rm h,1/2}$. 
Next we choose a value for $r_{\rm h,0}$ which can be used with $r_{\rm h,1/2}$ 
in eq.~(\ref{e:radevn}) to calculate an approximate $\zeta$ for the 
$1 > M/M_0 > 0.5$ phase of evolution. 
Eq.~(\ref{e:trh5ch}) with $N_0 \simeq 2 M_0 / \Msun$ 
(assuming the average stellar mass at $t = 0.0$ is roughly $0.5 \Msun$, 
which is true for the KTG3 IMF) 
gives $t_{\rm rh,0}$ which we use in eq.~(\ref{e:mtmngt}) to calculate 
$t_{1/2}$, the time taken for the cluster mass to reduce to half its 
initial value. 
From then on the initial parameters in eqs.~(\ref{e:radevn}) and 
(\ref{e:mtmngt}) can be replaced by the corresponding values at $t_{1/2}$ 
to find $M$ and $r_{\rm h}$ as a function of time with $\zeta = 5/3$. 
By iterating on this method we can find the initial values corresponding to 
current observed cluster properties. 
It should be noted that the solution is not single-valued:  
increasing $M_0$ and decreasing $r_{\rm h,0}$ gives similar values. 

There is a problem with this method for M67, demonstrated if we put 
$M = 2\,500 \Msun$ in eq.~(\ref{e:r2tide}) for the standard Galactic 
tidal field. 
This gives a tidal radius of $17.5\,$pc, corresponding to $r_{\rm h} = 4.4\,$pc 
which is almost twice what is observed. 
Francic~(1989) estimates a lower limit of $9\,$pc for the tidal radius of M67 
and finds no stars with high membership probabilities outside this range. 
Additionally the data of Fan et al.~(1996), from which the current mass is 
estimated, extend no further than $10\,$pc from the cluster centre. 
This agrees well with the observed half-mass radius of $2.5\,$pc and 
$r_{\rm t} = 4 r_{\rm h}$ from the model data. 
So does M67 contain less mass than is observed or does its tidal radius not 
correspond to the standard Galactic tidal field? 
Chernoff \& Weinberg~(1990) find that taking $v_{\rm\SSS G} = 220 \, {\rm km} 
\, {\rm s}^{-1}$ for $3 < R_{\rm\SSS G} < 20\,$kpc is consistent with current 
theoretical models of the mass distribution of the Galaxy. 
The position of M67 relative to the Sun gives $R_{\rm\SSS G} \simeq 9\,$kpc 
so the local tidal field should apply. 
However, the Galactic orbit of M67 is slightly eccentric 
(Carraro \& Chiosi~1994), with an apogalacticon of $9.09\,$kpc and a 
perigalacticon of $6.83\,$kpc, so it has been subject to a time varying 
tidal field.  
Possibly the structure of M67 was altered by an event in its past, 
such as an interaction with another cluster or an interstellar cloud 
(Terlevich 1987). 
We do not dwell on this non-standard tidal radius, preferring to discuss it 
further in Section~\ref{s:m67dis}. 
What is important is that the mass assumed for M67 corresponds to the 
observations from which it is derived, i.e. $2\,500 \Msun$ within $10\,$pc. 

Using this method we estimate that M67 had $N_0 \simeq 40\,000$ and 
$r_{\rm h,0} \simeq 1\,$pc to evolve to its current observed parameters. 
This is unfortunate because a simulation with $N_0 > 20\,000$ takes a 
prohibitively long time with currently available equipment,  
especially when using a 50\% binary fraction. 
On the other hand the main interest of this work is $N_{\rm\SSS BS}$ 
near $4\,200\,$Myr so it should be possible to extract meaningful results 
using a semi-direct method. 

\begin{table}
\begin{center}
\begin{minipage}{8cm}
\begin{tabular}{l|cccc|} \hline\hline
purpose & escape & escape & escape & M67 \\ 
$N$ & $10\,000$ & $10\,000$ & $1\,000$ & $15\,000$ \\ 
$t_0$ (Myr) & 0.0 & 0.0 & 0.0 & 2500.0 \\ 
distribution & & & & \\ 
model & Plummer & Plummer & Plummer & King, $W_0 = 7$ \\ 
$r_{\rm h,0}/r_{\rm t,0}$ & 0.1 & 
0.1\footnote{0.06 in one case} & 0.03 & 0.2 \\ 
$v_{\rm\SSS G}$ (${\rm km} \, {\rm s}^{-1}$) & 220 & 220 & 220 & 350 \\ 
$f_{\rm b}$ & 0.0 & 0.11 & 0.5 & 0.5 \\
binary & & & & \\ 
separations & EFT30 & EFT30 & EFT30 & EFT10, max 50 \\ 
\#sim & 3 & 3 & 6 & 1 \\ 
\hline\hline
\end{tabular}
\end{minipage}
\end{center}
\caption{
Model parameters for all simulations described in this work. 
The final row lists the number of times each simulation was performed.  
All simulations share the following general properties (see text for 
details): 
(i) single star masses taken from KTG3 IMF with minimum and 
    maximum mass limits of 0.1 and $50 \Msun$; 
(ii) binary masses taken from KTG1 IMF with a uniform distribution in $q$; 
(iii) thermal eccentricity distribution; 
(iv) $Z = 0.02$; and 
(v) escape radius of $2 r_{\rm t}$.  
}
\label{t:simpar}
\end{table}

\section{M67 $N$-body Model}
\label{s:m67nb4}

We start the simulation after the self-similar phase has begun, 
using an initial model that reflects the preceding stellar, binary and 
cluster evolution.
The model is substantially smaller in population number than the 
$N \simeq 40\,000$ required to start from the birth of the cluster, so it can 
be evolved directly to the age of M67 in a reasonable time. 
In view of the discussion of Section~\ref{s:nb4cde} and the small number 
statistics of the BS population, this is preferable to scaling up the results 
of a smaller simulation starting from initial conditions. 
For a first attempt using this method we begin a simulation with 
$N = 15\,000$ stars at $2\,500\,$Myr. 
The initial parameters for this semi-direct M67 model and for the 
simulations from which the escape rates were derived (see previous 
Section) are summarized in Table~\ref{t:simpar}. 

To estimate the binary fraction and the maximum 
orbital separation for the binaries in this starting model we consider 
the hard binary limit for the cluster. 
If indeed $N_0 = 40\,000$ and $r_{\rm h,0} = 1\,$pc then $a_{\rm hard} 
\simeq 10\,$AU for M67 initially, whereas it is now more like $140\,$AU. 
Only binaries which were initially hard could survive the early phases 
of cluster evolution so a conservative estimate of $50\,$AU is used for the  
upper limit of the separation distribution in the starting model. 
Lowering this limit would increase the incidence of interaction within 
binary systems and presumably lead to a larger number of BSs 
produced by the simulation. 
This would not be accompanied by an increase in the number of BSs  
found in wide eccentric orbits as the population of 
wide binaries available for exchange interactions is decreased. 
As already mentioned, observations of M67 indicate that the current binary 
fraction could be as high as 50\%. 
During the cluster evolution soft binaries are broken-up in dynamical 
encounters so this is a lower limit on the primordial fraction. 
However, hard binaries are retained preferentially by the cluster because 
they have a higher average mass than single stars. 
We therefore assume that the binary fraction has remained roughly constant 
during the cluster lifetime, in good agreement with the $N = 10\,000$ data, 
so we use $f_{\rm b} = 0.5$ for the starting model. 

\begin{figure}
\psfig{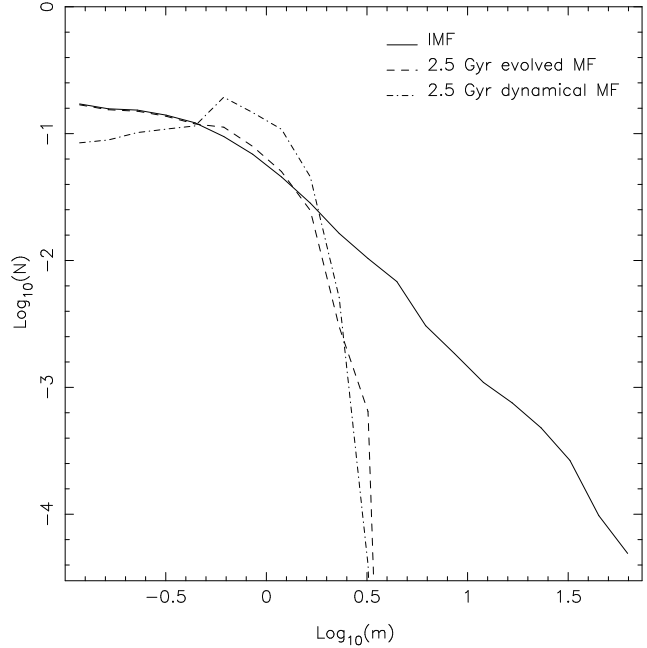}
\caption{
The full line shows the normalized IMF for a population with 11\% binaries 
where the single star masses are chosen from the KTG3 IMF and the binary masses 
from the KTG1 IMF. 
The dashed line shows the mass function (MF) of the same population evolved to 
$2\,500\,$Myr 
(note this is hidden by the full line for $\log m / \Msun < -0.35$) 
and the dash-dot line shows the MF of the same population 
evolved to $2\,500\,$Myr in the $N$-body code with a standard tidal field. 
}
\label{f:fig5bs}
\end{figure}

We use results from the previous simulations to estimate what the mass 
function (MF) will look like at $2\,500\,$Myr. 
Figure~\ref{f:fig5bs} shows the IMF for the $N = 10\,000$ models that included 
binaries and the corresponding MF for the population at $2\,500\,$Myr 
according to the population synthesis code and 
from the $N$-body simulations at the same age. 
It is evident that dynamics and the tidal field alter the MF, 
lowering it at the low-mass end and increasing the relative number of 
more massive stars. 
The number of BSs produced will be sensitive to the shape 
of the MF assumed for the starting model as a larger proportion of 
systems with mass comparable to that of the MS turn-off at a particular  
time will lead to more stragglers at that time. 
To generate the starting population for the semi-direct M67 simulation, we  
evolve a large population with a 50\% binary fraction and $Z = 0.02$ to an 
age of $2\,500\,$Myr using the population synthesis code alone. 
The single star masses are chosen from the KTG3 IMF and the binary masses 
and parameters are chosen in the same way as for the population 
synthesis run PS6, but with an upper limit of $50\,$AU in the separation 
distribution.  
Then using the information gained from the dynamically altered MF in 
Figure~\ref{f:fig5bs}, i.e. the ratio of stars in each mass bin between the 
dynamical and non-dynamical MFs, we take $5\,000$ single stars and $5\,000$ 
binaries from the large population to populate the starting model 
(see Table~\ref{t:mftab}). 
This gives a starting mass of $6\,000 \Msun$ for the cluster at $2\,500\,$Myr. 

\begin{figure}
\psfig{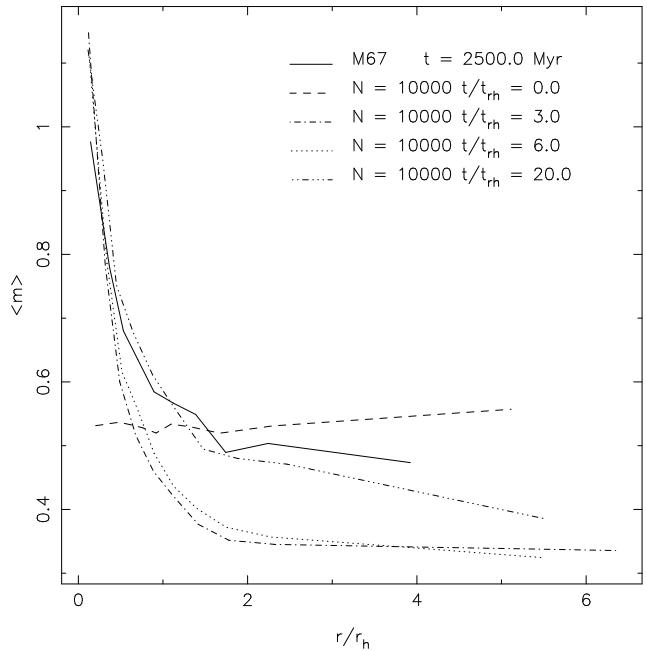}
\caption{
Profile of the average stellar mass in successive Lagrangian shells. 
The radius is scaled by the cluster half-mass radius. 
Each Lagrangian shell contains 10\% of the cluster mass. 
The profiles plotted are for the combined $N = 10\,000$ models data at 
$t/t_{\rm rh}$ = 0.0, 3.0 (near core-collapse), 6.0 (when the model has lost 
half its mass) and 20.0, as well as for the initial M67 model at $2\,500\,$Myr. 
}
\label{f:fig6bs}
\end{figure}

\begin{table}
\setlength{\tabcolsep}{0.2cm}
\begin{center}
\begin{tabular}{cccrr} \hline\hline
lower & upper & number   & mass in$\,\,\,\,$ & mass in  \\
mass  & mass  & fraction & single stars      & binaries \\
\hline
0.100 & 0.137 & 0.081 &  94.8 &    0.0 \\
0.138 & 0.190 & 0.089 & 144.7 &    0.0 \\
0.191 & 0.263 & 0.097 & 170.5 &   50.4 \\
0.264 & 0.363 & 0.105 & 199.5 &  124.3 \\
0.364 & 0.502 & 0.107 & 220.7 &  236.9 \\
0.503 & 0.694 & 0.193 & 362.3 &  785.6 \\
0.695 & 0.960 & 0.148 & 378.3 &  827.6 \\
0.961 & 1.326 & 0.116 & 272.1 & 1038.1 \\
1.327 & 1.833 & 0.055 & 118.6 &  732.5 \\
1.834 & 2.532 & 0.005 &  16.9 &   83.6 \\
2.533 & 3.500 & 0.004 &   0.0 &   98.6 \\
\hline
\end{tabular}
\end{center}
\caption{
Mass groups used to populate the M67 starting model. 
All masses are in units of $\Msun$. 
The lower and upper mass limits of each group are given in the first 
two columns, followed by the number fraction of stars in that group for 
the combined $N = 10\,000$ with $f_{\rm b} = 0.11$ models at $2\,500\,$Myr. 
$5\,000$ single stars and $5\,000$ binaries, all evolved to an age of 
$2\,500\,$Myr (see text for details), 
are chosen for the M67 starting model according to these number fractions. 
The resulting masses in single stars and in binaries for each mass group 
in the M67 starting model are given in the final two columns. 
}
\label{t:mftab}
\end{table}

We use a tidal field with $v_{\rm\SSS G} = 350 \, {\rm km} \, {\rm s}^{-1}$ 
at $R_{\rm\SSS G} = 8.5\,$kpc which fixes $r_{\rm t} = 17\,$pc.  
This tidal radius determines the length scale used in the $N$-body 
simulation. 
The stars are distributed according to a multi-mass King model 
(King 1966; Chernoff \& Weinberg 1990) with $W_0 = 7$ and a central number 
density $n_0 = 2.28 \times 10^3 \, {\rm pc}^{-3}$. 
The concentration of the model is determined by the dimensionless 
parameter 
\beq
W_0 = \frac{{\Psi}_0}{{\sigma}^2} \, , 
\eeq 
where ${\Psi}_0$ is the central potential and we use a central velocity 
dispersion ${\sigma}^2 = 3 \times 10^{10} \, {\rm cm}^2 \, {\rm s}^{-2}$. 
These positions are scaled so that the cluster just fills the tidal radius,  
which results in a half-mass radius of $3.4\,$pc for the starting model. 
Figure~\ref{f:fig6bs} shows the evolution of the average stellar mass profile 
for the combined $N = 10\,000$ model data.  
It can be seen that as the models evolve, two-body effects cause the heavier 
stars to segregate towards the inner regions. 
Using the King model to determine the initial spatial distribution of the stars 
for the M67 simulation builds in a degree of mass-segregation 
so that this energy equipartition is taken into account. 

We can estimate the relation between the mass of our starting model and its 
mass at $4\,200\,$Myr using the escape rates discussed in 
Section~\ref{s:m67esc}, but this depends on how binaries contribute to the 
value of $N$ used in the calculation.  
If we assume that relatively hard binaries act as single stars 
when modelling relaxation effects then $N$ would be $10\,000$ and the mass at 
$4\,200\,$Myr should be about $1\,500 \Msun$. 
On the other hand, if the binary components behave as single stars then 
$N = 15\,000$ and the mass left would be about $3\,000 \Msun$. 
Either way this is close enough to the mass derived from observations to 
proceed with the simulation. 

We evolve the model to $T = 4\,310\,$Myr using {\tt NBODY4}. 
At this time the cluster mass is $1\,140 \Msun$. 
It consists of 560 single stars and 740 binaries, 
the tidal radius is $10\,$pc and the half-mass radius is $2.5\,$pc. 
The mass in single stars and binaries with masses greater than $0.5 \Msun$ is 
$1\,050 \Msun$, more than 90\% of the total mass. 
The simulation lasted for $1\,260$ $N$-body time units and took one month  
dedicated use of the HARP-3.  

\begin{figure}
\psfig{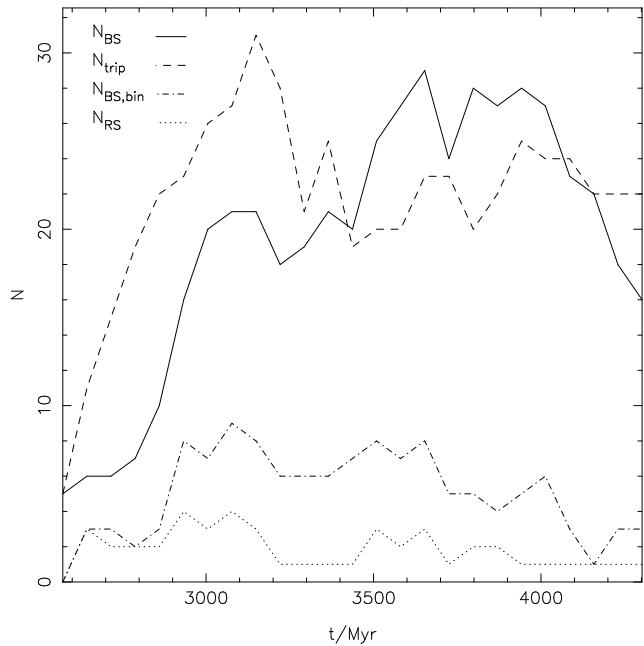}
\caption{
The evolution of the number of blue stragglers (full line), triple 
systems (dashed line), blue straggler binaries (dash-dot line), and 
RS$\,$CVn systems (dotted line), in the M67 simulation. 
}
\label{f:fig7bs}
\end{figure}

\begin{figure}
\psfig{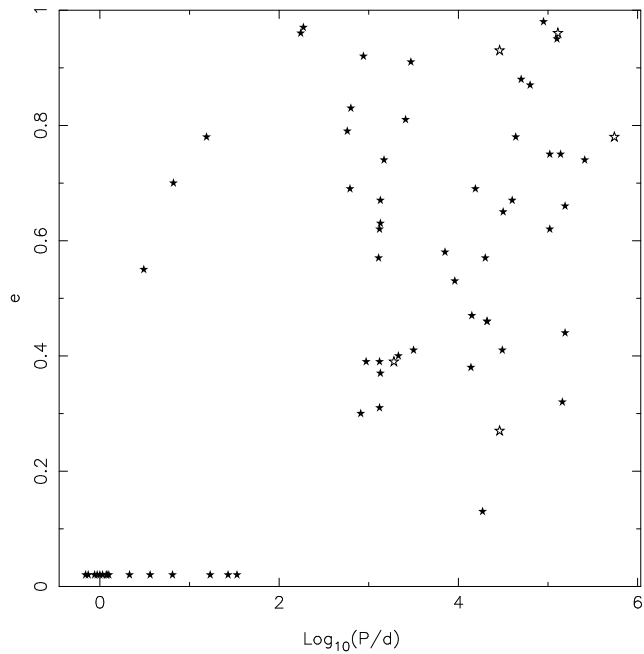}
\caption{
Distribution of periods and eccentricities for binaries formed 
with a blue straggler component during the M67 simulation. 
The open points represent BS-BS binaries, i.e. each component is a BS. 
}
\label{f:fig8bs}
\end{figure}

Figure~\ref{f:fig7bs} shows the evolution of the number of BSs 
present in the simulation. 
Also shown are the numbers of triple systems, RS$\,$CVn systems and binaries 
containing a BS. 
At $T = 4\,200\,$Myr there are 22 BSs in the model but only one 
of these is in a binary. 
There is only one RS$\,$CVn system in the cluster at this time. 
The highest number of BSs present at any one time is 29 at 
$T = 3\,653\,$Myr with seven of these in binaries. 
At this time there are three RS$\,$CVn systems. 

\begin{figure*}
\psfig{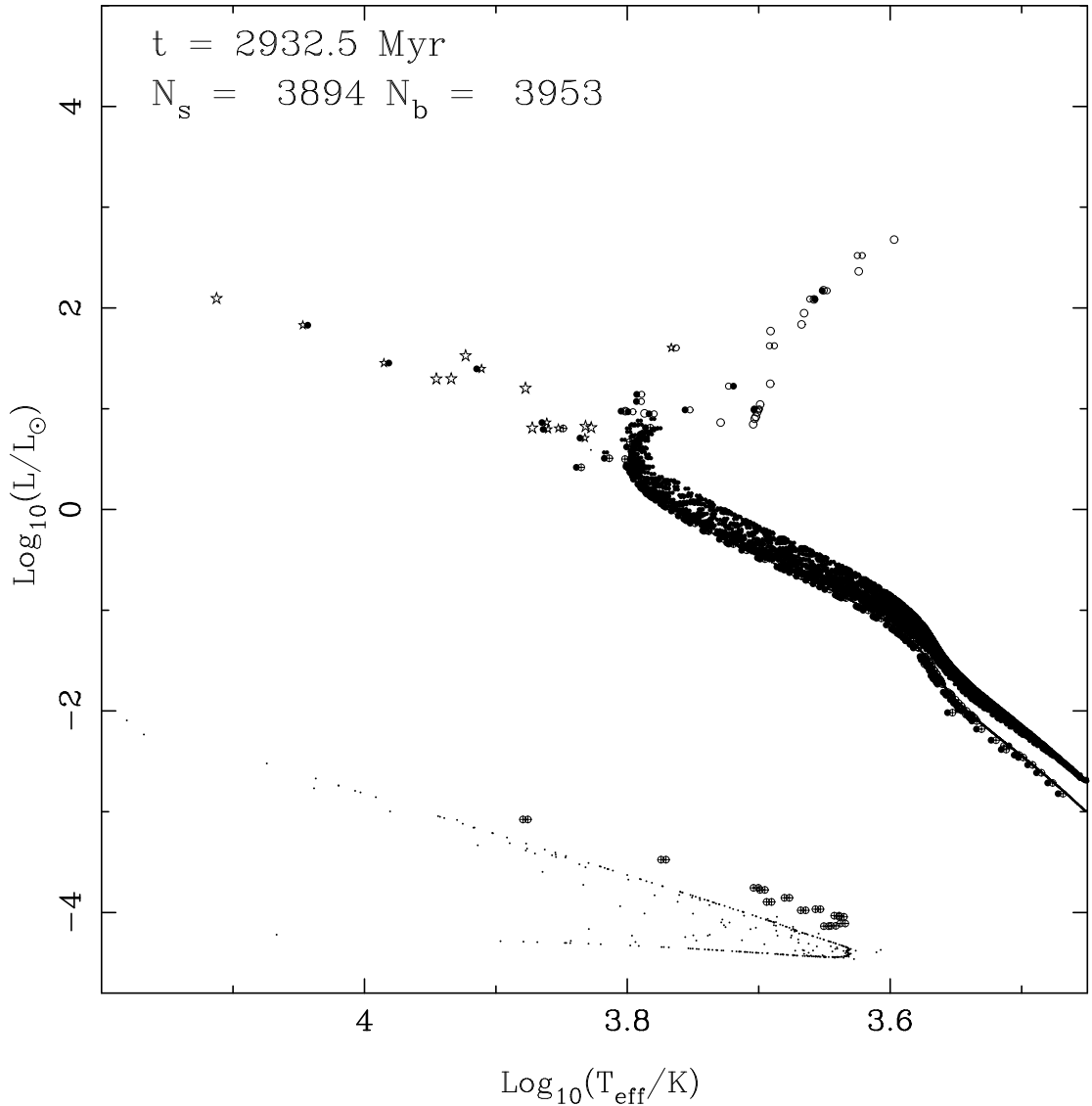}
\caption{
Hertzsprung-Russell diagram for the M67 $N$-body simulation at an age of 
$2\,932.5\,$Myr when $3\,894$ single stars and $3\,953$ binaries remain. 
Main-sequence stars (dots), blue stragglers (open stars), 
sub-giants, giants and naked helium stars (open circles) 
and white dwarfs (dots) are distinguished. 
Binary stars are denoted by overlapping symbols appropriate 
to the stellar type of the components, with main-sequence binary components 
depicted with filled circles and white dwarf binary components as 
$\oplus$ symbols.
The effective temperature of a binary is computed according to 
Hurley \& Tout~(1998). 
}
\label{f:fig1hr}
\end{figure*}

\begin{figure*}
\psfig{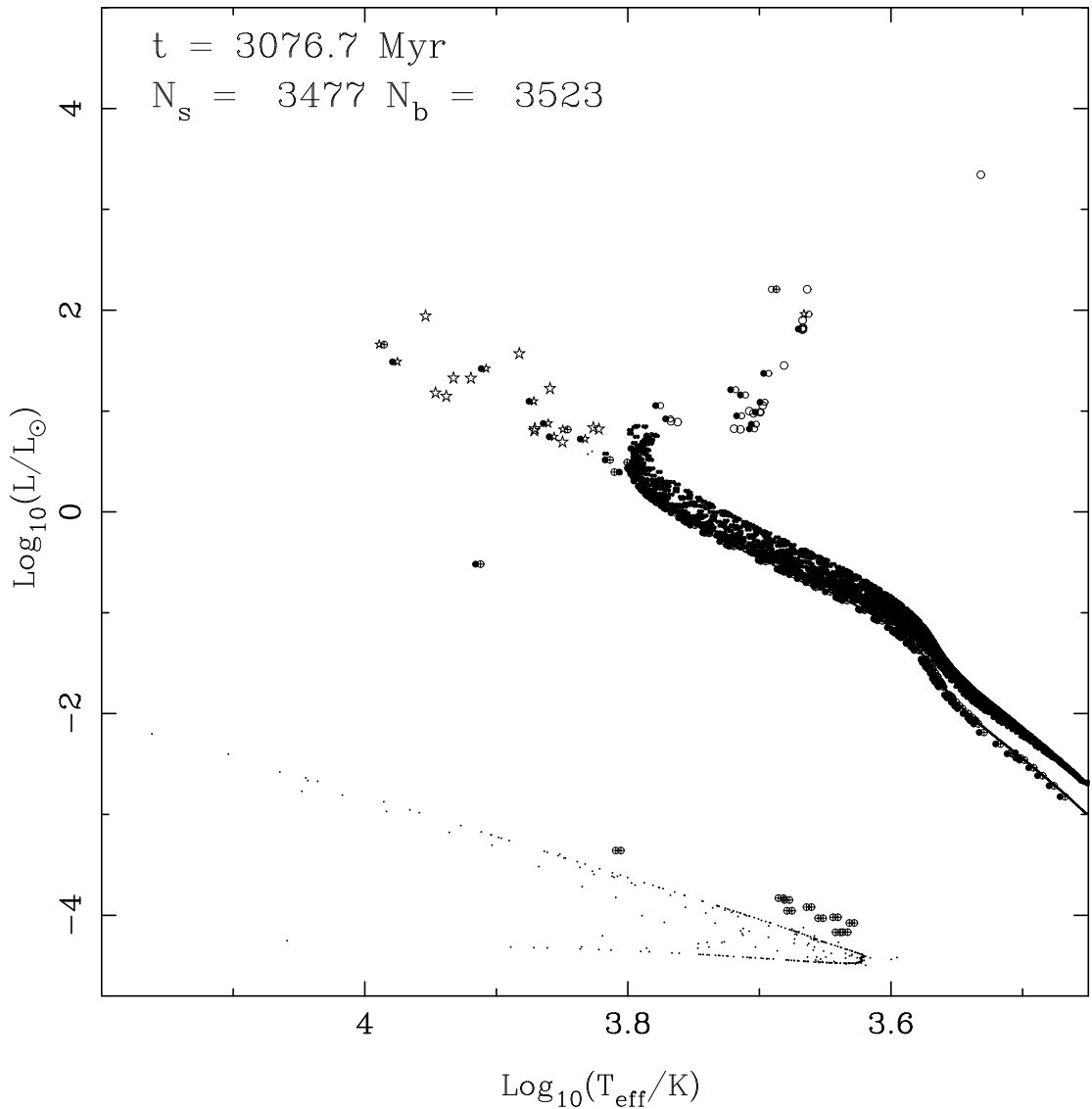}
\caption{
As Figure~\ref{f:fig1hr} at $3\,076.7\,$Myr. 
Present in the cluster at this time are 21 blue stragglers, nine of which are  
in binaries. 
One GB star has a BS binary companion in an eccentric orbit. 
A CHeB star and a WD in an eccentric binary is also present. 
There are ten cataclysmic variables and ten double-degenerate systems. 
}
\label{f:fig2hr}
\end{figure*}

\begin{figure*}
\psfig{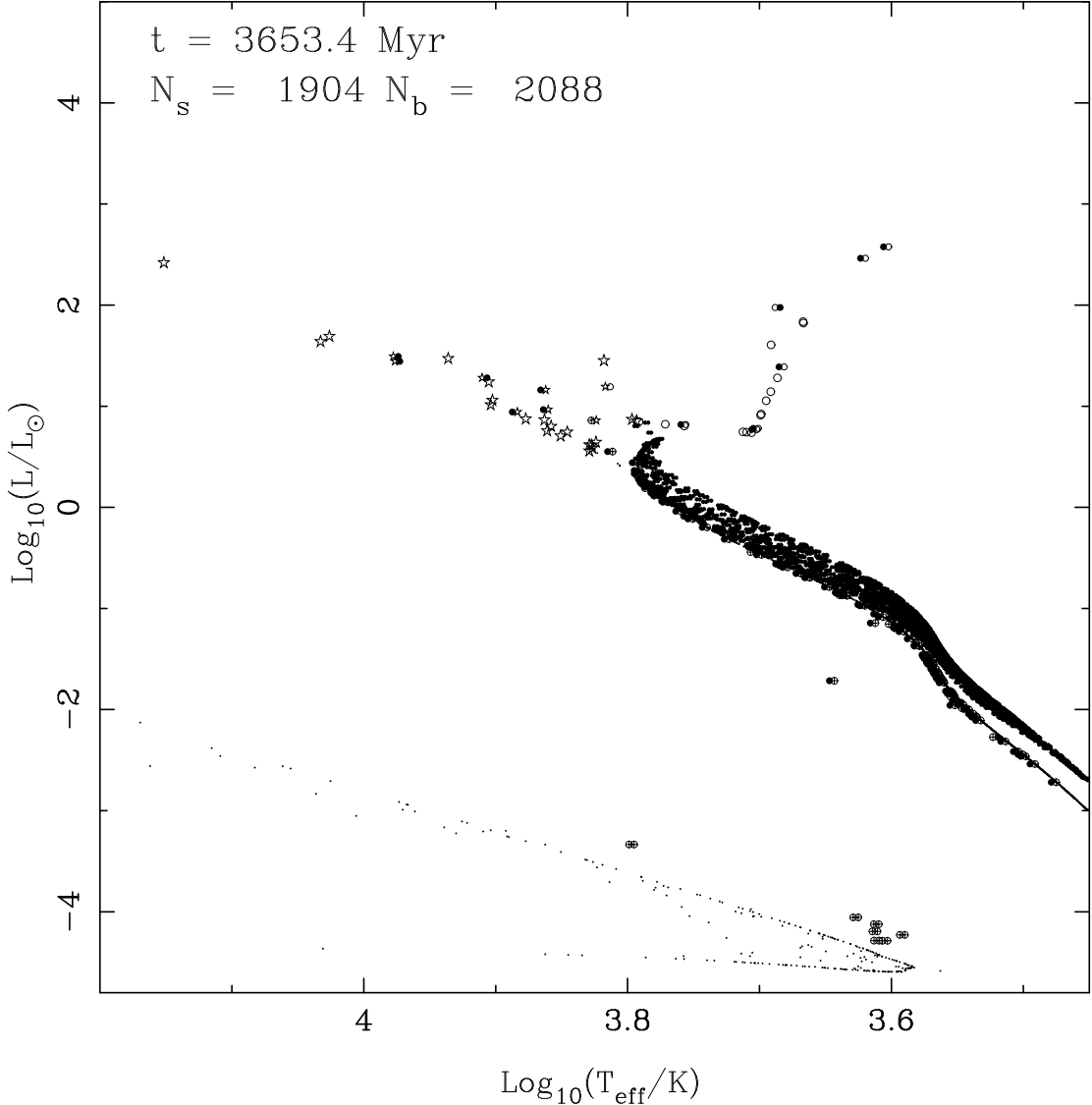}
\caption{
As Figure~\ref{f:fig1hr} at $3\,653.4\,$Myr.  
Present in the cluster at this time are 29 blue stragglers, the maximum 
number during the simulation, with eight of these in binaries. 
There is one super-BS, five cataclysmic variables and 
seven double-degenerate systems. 
The depletion of double-degenerate systems is due to ejection of some close 
systems and break-up of some wide systems. 
}
\label{f:fig3hr}
\end{figure*}

\begin{figure*}
\psfig{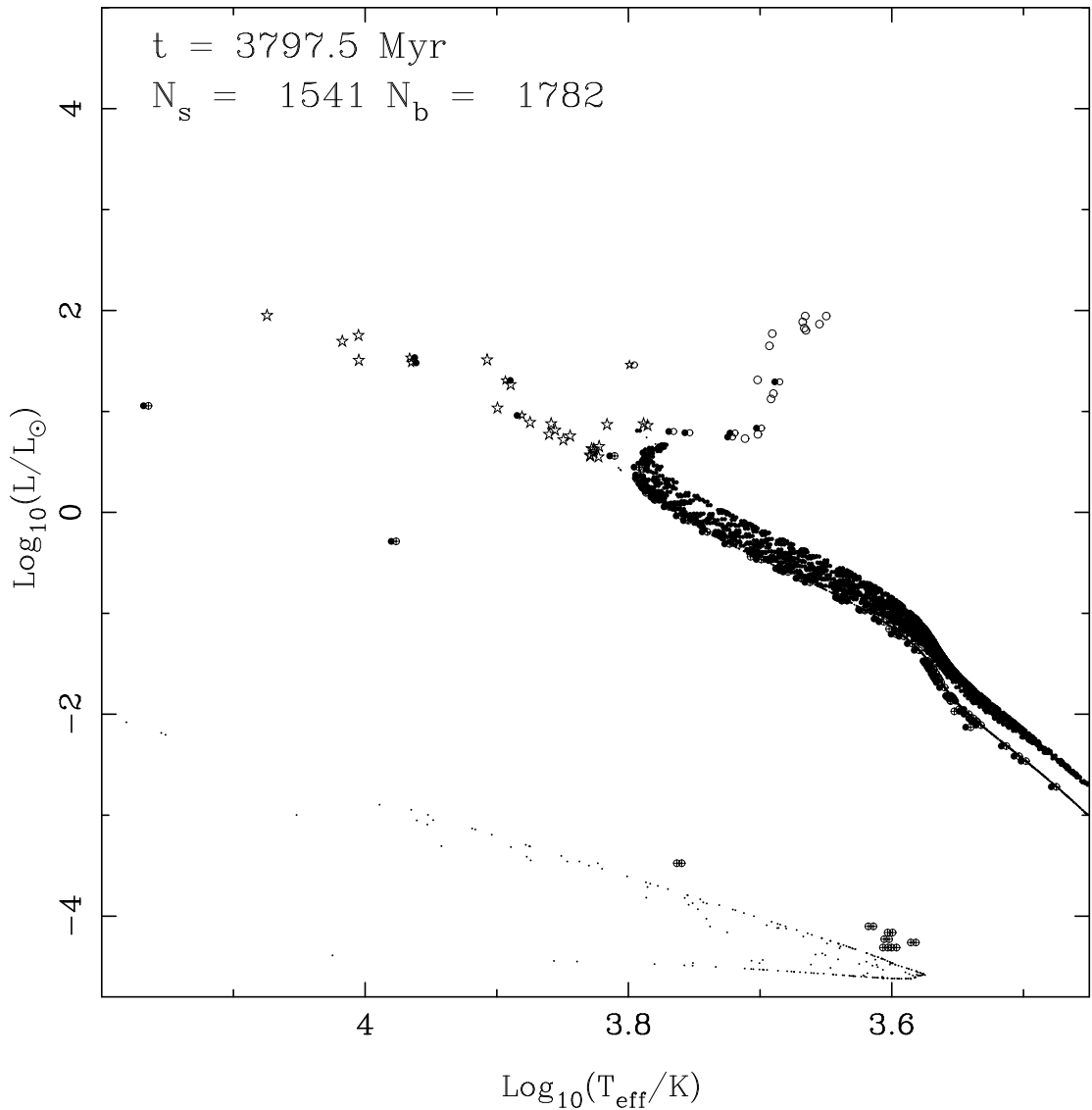}
\caption{
As Figure~\ref{f:fig1hr} at $3\,797.5\,$Myr. 
Present in the cluster at this time are 28 blue stragglers with five of these 
in binaries. 
One BS binary has a GB star companion which is currently filling its 
Roche-lobe and transferring mass. 
There is one super-BS, seven cataclysmic variables and 
seven double-degenerate systems. 
}
\label{f:fig4hr}
\end{figure*}

\begin{figure*}
\psfig{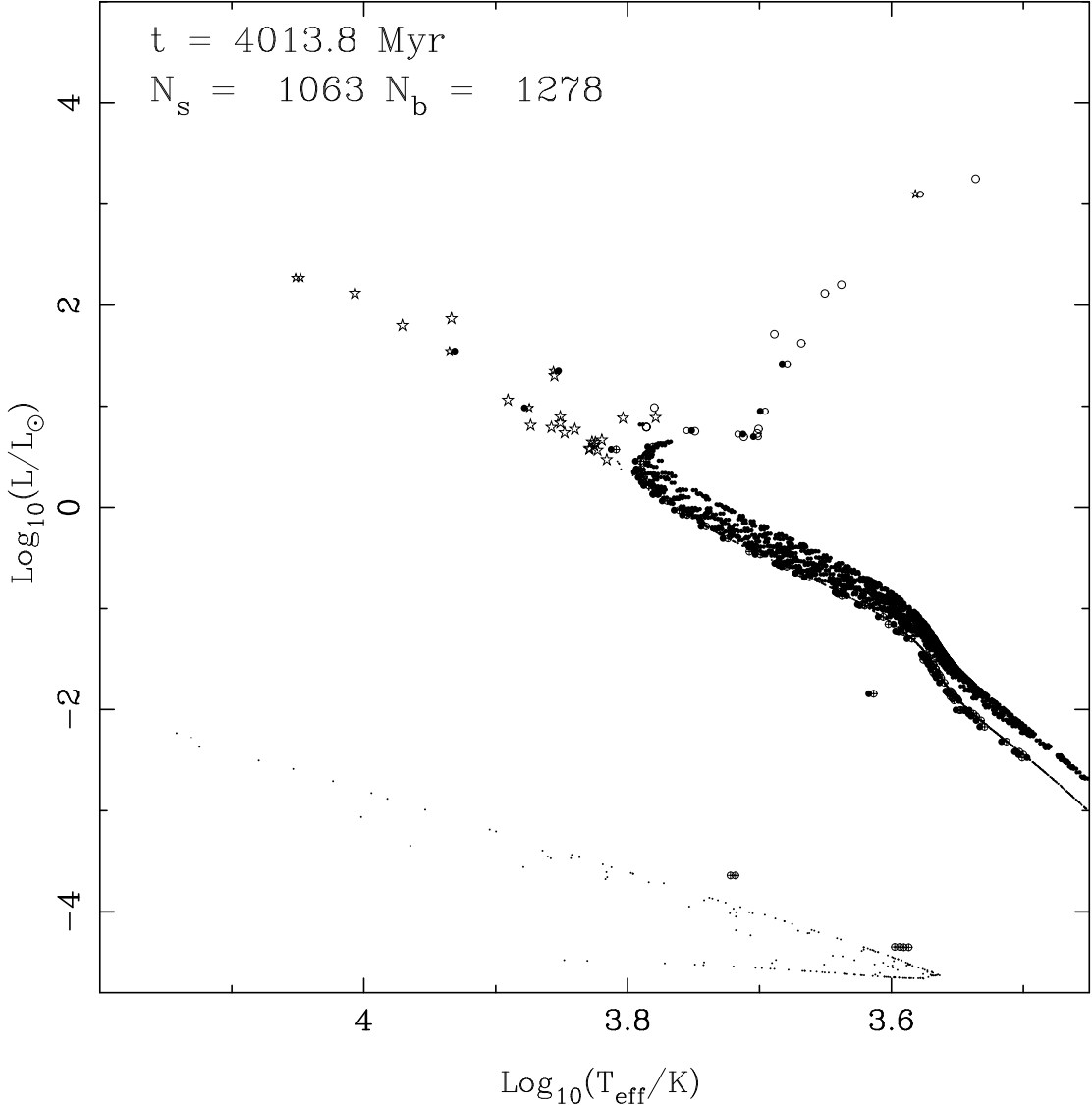}
\caption{
As Figure~\ref{f:fig1hr} at $4\,013.8\,$Myr. 
Present in the cluster at this time are 27 blue stragglers with six of these 
in binaries. 
There are two super-BSs and one BS-BS binary. 
There are nine cataclysmic variables and three double-degenerate systems. 
}
\label{f:fig5hr}
\end{figure*}

\begin{figure*}
\psfig{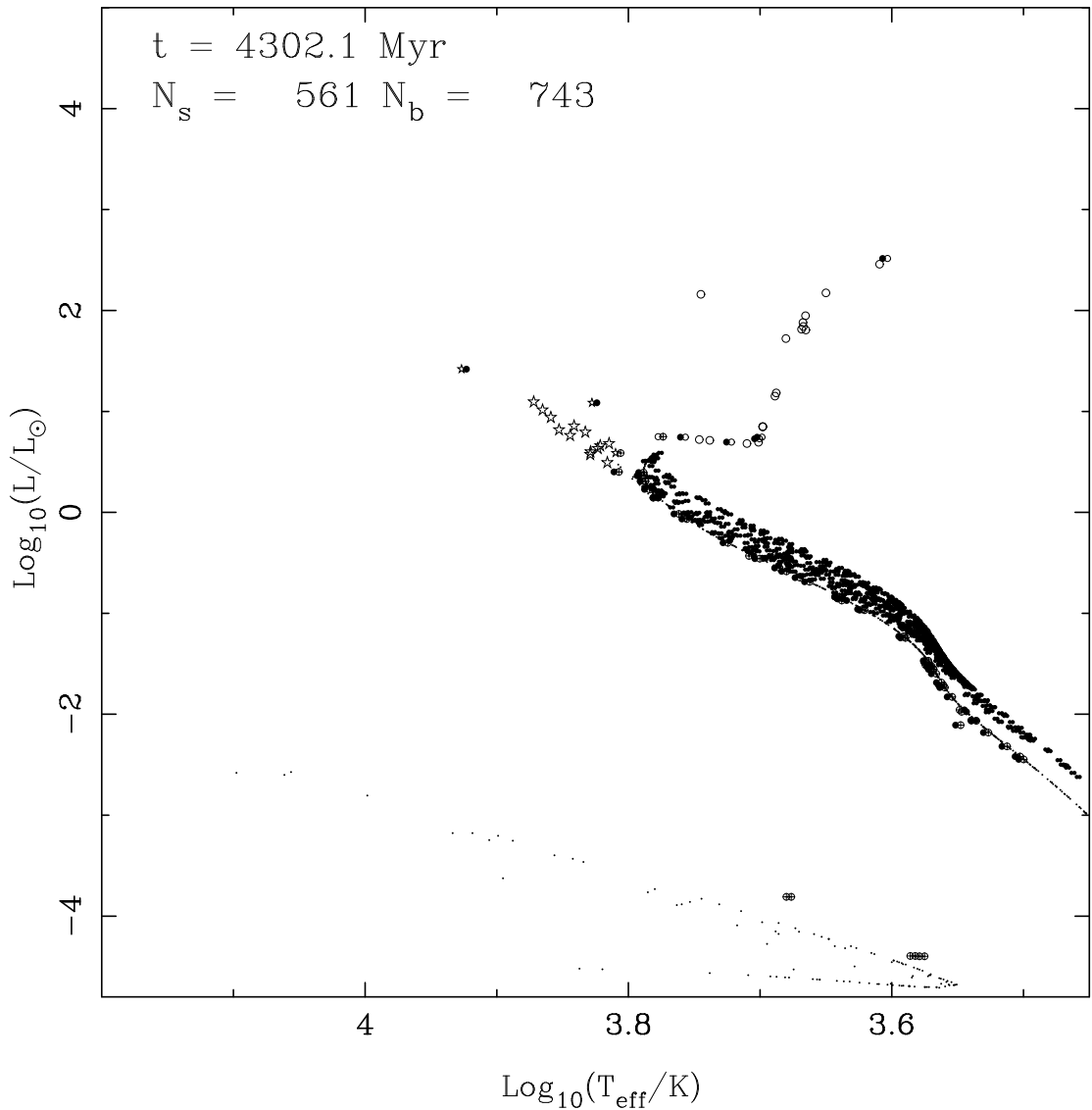}
\caption{
As Figure~\ref{f:fig1hr} at $4\,302.1\,$Myr (the end of the simulation). 
Present in the cluster at this time are 16 blue stragglers with three of these 
in binaries. 
There are six cataclysmic variables and three double-degenerate systems. 
}
\label{f:fig6hr}
\end{figure*}

\begin{figure*}
\psfig{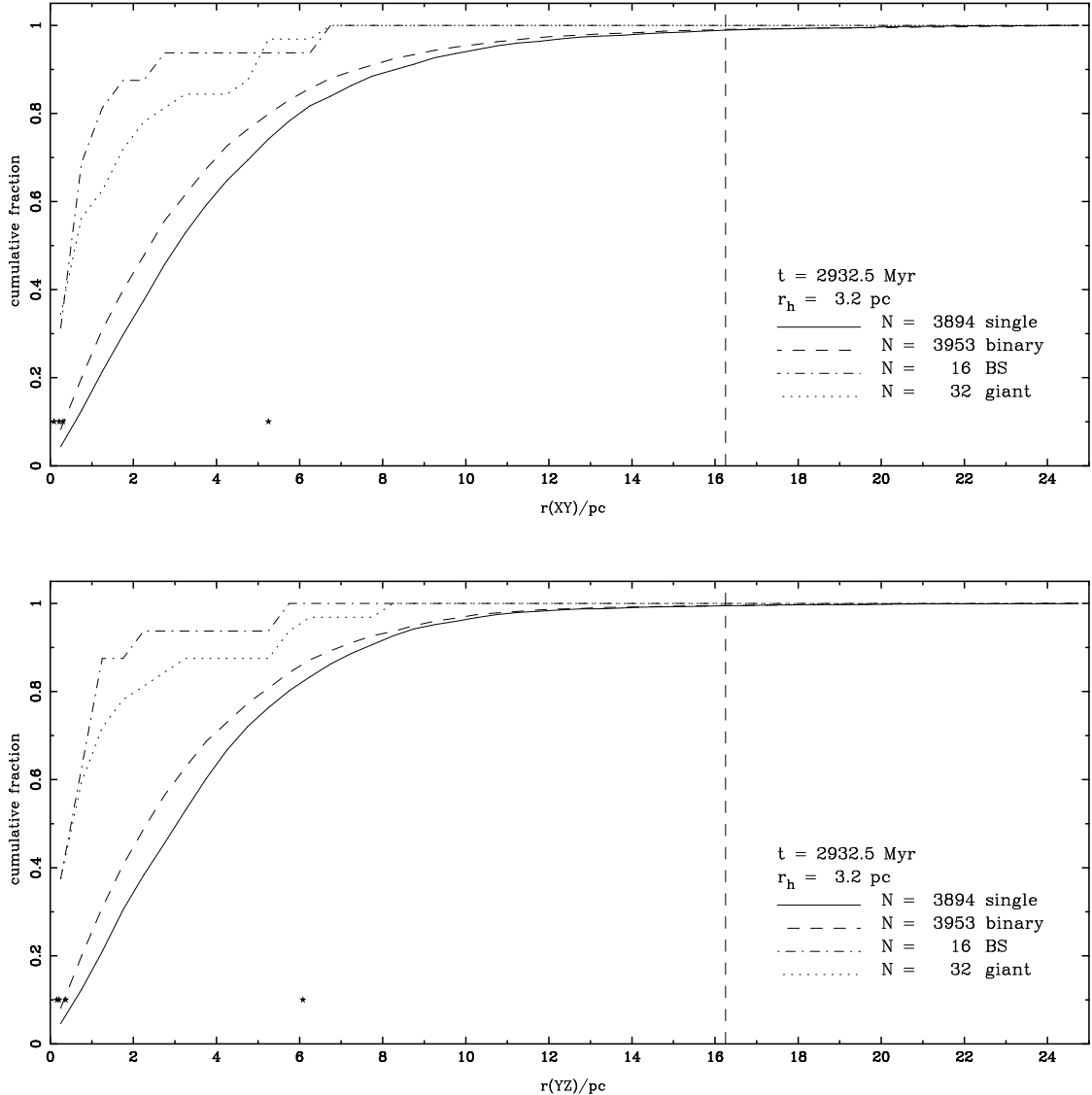}
\caption{
Cumulative radial profiles in the XY- and YZ-planes for the population shown in 
Figure~\ref{f:fig1hr}. 
The tidal radius is $16.3\,$pc (vertical dashed line), 
the half-mass radius is $3.2\,$pc and the 
half-mass radius of the blue straggler stars is $0.8\,$pc. 
Note that stars are not actually removed from the simulation until they are at 
a distance greater than two tidal radii from the cluster centre.  
Both blue stragglers and giants congregate towards the centre. 
The four RS$\,$CVn systems present are plotted as filled stars. 
}
\label{f:fig1xy}
\end{figure*}

\begin{figure*}
\psfig{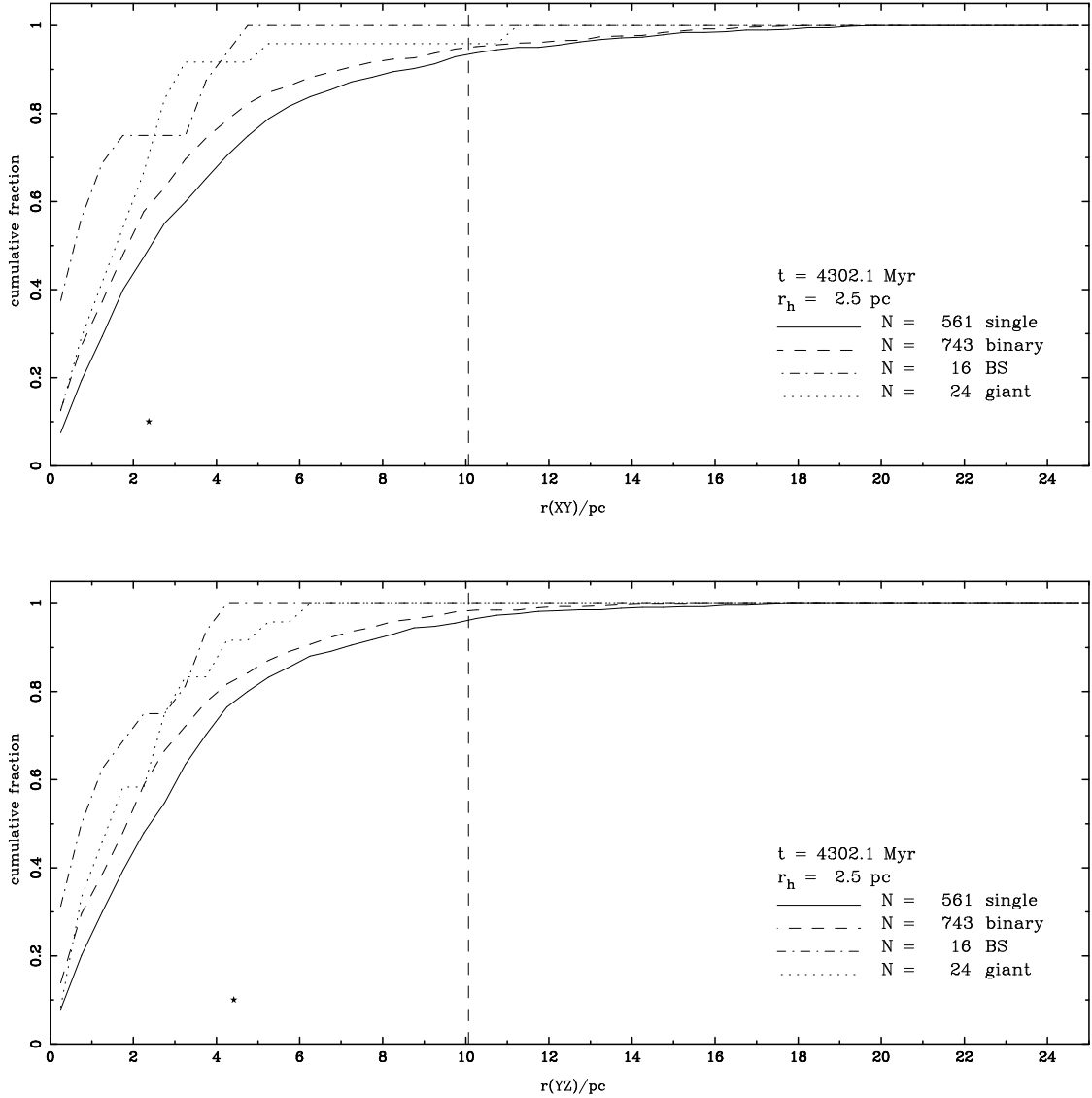}
\caption{
Cumulative radial profiles in the XY- and YZ-planes for the population shown in 
Figure~\ref{f:fig6hr}. 
The tidal radius is 10.1~pc (vertical dashed line) and the blue straggler 
half-mass radius is $0.9\,$pc. 
The RS$\,$CVn system present is plotted as a filled star. 
}
\label{f:fig6xy}
\end{figure*}

Shown in Figure~\ref{f:fig8bs} is the period-eccentricity distribution of all 
BS binaries formed during the simulation. 
Of these, five are BS-BS systems. 
There appears to be two fairly distinct BS binary populations, one consisting  
of close circular orbits and the other with wider eccentric orbits. 
The BSs in circular orbits formed by stable mass-transfer, 
beginning in general when the primary is in the Hertzsprung gap (HG).  
These still have their primordial companion which evolves to become a WD. 
No instance of a BS formed from Case~C mass transfer, or as a 
result of wind accretion, occurs in this simulation. 
For the relatively wide binaries, $P > 100\,$d, the distribution appears 
uniform for $e > 0.2$. 
These binaries form in hierarchical systems or as a result of an exchange 
interaction. 
A star is more likely to be exchanged into a wide orbit than into an 
existing hard binary.  
We also expect that the eccentricities of newly formed binaries follow a 
thermal distribution (Heggie 1975), which is proportional to the eccentricity.  
Furthermore, since exchange interactions tend to take place in the 
dense central regions of the cluster it is not surprising to find a 
paucity of wide circular binaries. 
A total of 53 exchange interactions were recorded during the simulation. 
Of these, 18 are the result of a direct exchange and the remaining 35 are the 
result of 12 resonant exchanges (Heggie 1975). 
In a resonant exchange the three (or four) stars involved in the interaction 
form a temporary bound system which can then undergo a series of exchanges 
in quick succession before one (or two) of the stars escapes. 

On average half the BSs formed during the simulation would 
have done so without the introduction of cluster dynamics. 
We note however that the cluster environment was taken into account when 
choosing the binary parameters for the starting model, effectively 
doubling the number of BSs expected when compared with population 
synthesis run PS6. 
Consider the 16 BSs present when the simulation ended, of which 
three are in binaries. 
Seven of these formed from standard Case~A mass transfer in a primordial 
binary: the two components coalesced to leave a single BS. 
One formed as a result of standard Case~B mass transfer and is in a circular 
orbit with a $17\,$d period with its original companion, now a WD. 
Another two formed from Case~A mass transfer but only after a series of 
close encounters had altered the orbital parameters and caused each system to 
circularize. 
Without perturbations to their orbits these stars would not have been close 
enough to interact. 
The remaining BSs are the result of collisions within binary systems, three 
after a star was exchanged into a highly eccentric orbit and the other 
three after perturbations to the existing binary increased the eccentricity 
and caused the orbit to become chaotic. 
One of the BS binaries that formed from a dynamical interaction has a 
period of $813\,$d and $e = 0.3$ and the other has $P = 6.6\,$d and $e = 0.7$. 
The BS in the wider of these has a mass $M_1 = 1.6 \Msun$ and was created 
after Case~A mass transfer in a primordial binary lead to coalescence. 
Later in the cluster evolution it underwent an exchange interaction with a 
wide binary to form the existing binary. 
The companion is a MS star with $M_2 = 1.3 \Msun$ and the star that left the 
temporary three-body system was the lightest of the three, $M_3 = 0.7 \Msun$.  
The other binary was originally part of a three-body hierarchy 
consisting of an outer component in an eccentric orbit about a  
primordial binary with $P = 301\,$d and $e = 0.18$.  
Inclination-induced perturbations drove the inner eccentricity up to 0.94 at 
which point one of the inner MS components collided with the third MS star,  
producing the BS, which remains bound to its original companion. 

Figures~\ref{f:fig1hr}, \ref{f:fig2hr}, \ref{f:fig3hr}, \ref{f:fig4hr}, 
\ref{f:fig5hr} and \ref{f:fig6hr} show the Hertzsprung-Russell diagram 
of the cluster model at various epochs. 
The spatial distribution at $2\,933\,$Myr and at $4\,302\,$Myr is shown in 
Figures~\ref{f:fig1xy} and \ref{f:fig6xy} respectively, as the 
cumulative radial profiles in XY- and YZ-planes\footnote{
Full colour versions of the Hertzsprung-Russell diagrams and the 
spatial distributions, at a larger number of epochs, can be found at 
www.ast.cam.ac.uk/$\sim$cat/m67.html.}. 
It is evident that the BSs are concentrated towards the core 
of the cluster. 

At $T = 4\,014\,$Myr (Figure~\ref{f:fig5hr}) there are two super-BSs 
in the cluster. 
One of these has a mass 2.3 times the then turn-off mass of $1.32 \Msun$.  
The other has a mass of $2.7 M_{\rm TO}$ and is in a binary with another BS. 
A total of eight super-BSs formed during the simulation. 
Also present in the cluster at $T = 4\,014\,$Myr are nine cataclysmic 
variables and three double-degenerate systems. 
There are 19 giants, six of which are in binaries. 
One of the 27 BSs is in a binary with an AGB star separated by 
$4 \times 10^{-4}\,$pc. 
This is about $0.1''$ at the distance of M67, so it 
would not appear as a BS on a cluster CMD (the CCD used by MMJ had 
$0.77''$/pixel). 

The $2.7 M_{\rm TO}$ super-BS in the model at $4\,014\,$Myr is an 
interesting case. 
At the beginning of the simulation it is a single star with mass 
$M_1 = 1.33 \Msun$. 
At $3\,190\,$Myr it enters a triple system in which it 
exchanges into the original binary. 
The resulting binary has orbital parameters $P = 9\,120\,$d and $e = 0.52$ and 
survives to $3\,740\,$Myr when it is involved in a binary-binary encounter. 
This leaves the proto-BS in an orbit of $P = 153\,$d and $e = 0.99$ with 
a companion of mass $M_2 = 1.0 \Msun$. 
Owing to the large eccentricity the binary components collide and merge to 
form a BS of mass $M_1 = 2.33 \Msun$. 
At $3\,870\,$Myr this BS is involved in a three-body interaction and forms 
another binary, this time with a star of mass $M_2 = 2.28 \Msun$ which is 
itself a BS. 
The BS-BS binary has $P = 1\,350\,$d and $e = 0.67$. 
At $3\,990\,$Myr this binary forms a stable four-body hierarchy with yet 
another binary, 
the original BS collides with one of the stars in the other binary, 
forms the super-BS with $M_1 = 3.45 \Msun$, and remains bound to the other BS. 
The fourth body escapes and leaves the binary that is observed at 
$4\,014\,$Myr. 
Later at $4\,080\,$Myr the super-BS and BS collide when the eccentricity of the 
orbit has grown to 0.93 as a result of external perturbations. 
This makes a BS with $M_1 = 5.7 \Msun$, i.e. $4.4 M_{\rm TO}$.  
The new super-BS captures a companion at $4\,112\,$Myr, with which it then 
collides to form a super-BS with $M_1 = 7.7 \Msun \simeq 6 M_{\rm TO}$. 
Soon after it strongly interacts 
with a hard binary and gets ejected from the cluster.

\section{Discussion}
\label{s:m67dis}

An obvious problem with our M67 model is the lack of BSs in wide 
circular binaries because two such systems, S975 and S1195, are observed in 
the real cluster. 
Leonard~(1996) suggests these may be the result of mass transfer from a 
core helium-burning (CHeB) primary star.  
This does not seem likely because a star does not get any bigger during CHeB 
than it did on the GB, so if the primary does not fill its Roche-lobe on the 
GB, i.e. Case~B mass transfer, Case~C mass transfer is very unlikely before  
the star has reached the AGB. 
If mass transfer does occur when the primary is on the AGB then 
common-envelope evolution can only be avoided if the primary has  
lost enough of its envelope to become the less massive star. 
Wide binaries can form via exchange in the cluster but the results show this 
is unlikely to produce circular orbits. 
So it would seem that a number of wide circular BS binaries should be 
formed by isolated evolutionary processes, either from stable Case~C mass 
transfer or wind accretion. 
However, hardly any such binaries are produced with the period distributions 
used in the population synthesis runs. 
Perhaps this indicates that a bi-modal period distribution is required, or that 
the peak in the distribution should cover a wider range of periods.  
A more likely solution is that the wind-accretion efficiency we have used 
is too low.  
Our treatment of wind accretion in the binary evolution model assumes 
that the wind velocity $V_{\rm\SSS W}$ is simply $V_{\rm esc}/\sqrt{2}$ 
where $V_{\rm esc}$ is the escape velocity from the surface of the star 
(see Section~2.1 of Hurley, Tout \& Pols~2000 with $\beta = 0.5$). 
The accretion rate is approximately proportional to $V_{\rm\SSS W}^{-4}$ 
so that a small error in $V_{\rm\SSS W}$ has a large effect. 
We repeated run PS6 with the wind velocity reduced by a factor of 2, 
i.e. $\beta = 1/8$ in Section~2.1 of Hurley, Tout \& Pols~(2000).   
This is actually more in keeping with observations. 
The result is 
$N_{\rm\SSS BS} = 5.1$ and $N_{\rm\SSS RS} = 4.3$ at $4.2\,$Gyr 
with 23\% of the BSs in close binaries and 16\% in wide binaries ($P > 1\,$yr). 
The average period of these wide BS binaries is $2\,000\,$d which is consistent 
with the M67 BS binaries. 
In future $N$-body simulations we will incorporate this variation. 
Furthermore, if we use the weaker criterion of Webbink (1988) to determine the 
onset of dynamical mass transfer at RLOF (see Section~2.6.1 
of Hurley, Tout \& Pols~2000), an 
additional channel for the formation of binary BSs is created. 
Repeating run PS6 with $q_{\rm crit}$ according to Webbink (1988), 
together with $\beta = 1/8$, gives $N_{\rm\SSS BS} = 5.5$ at $4.2\,$Gyr 
with 22\% of the BSs in close binaries and 22\% in wide binaries. 

None of the eccentric BS binaries in the model are the result of stable mass 
transfer followed by perturbations to the circular orbit. 
Primarily this is because no BSs are formed in wide circular binaries in the 
first place, but even if they were the eccentricity induced would be small 
and tidal forces would quickly return the orbit to circularity. 
Rasio \& Heggie~(1995) show that the induced eccentricity $e_{\rm f}$ varies as
\beq
e_{\rm f} \propto \left( \frac{r_{\rm p}}{a} \right)^{-5/2} \, , 
\eeq
where $r_{\rm p}$ is the separation at periastron. 
Therefore an eccentric orbit is more likely to be altered by a close encounter. 
Assuming a stellar number density of $20 \, {\rm pc}^{-3}$ and a velocity 
dispersion of $0.40 \, {\rm km} \, {\rm s}^{-1}$ in the core of M67,  
Leonard~(1996) uses the result of Rasio \& Heggie~(1995) to derive a mean 
induced eccentricity of $e \simeq 10^{-3}$ for wide, $P = 10^4\,$d, 
circular orbits. 

Not surprisingly most of the eccentric BS binaries are the result of exchange 
interactions and none are produced from tidal capture. 
The exchange timescale can be expressed as 
\beq 
{\tau}_{\rm ex} = \frac{1}{n_{\rm b} \, \Sigma \, V_{\rm rel}} \, , 
\eeq
where $n_{\rm b}$ is the binary number density, $V_{\rm rel}$ is the relative 
speed of the third body and the binary centre-of-mass, and $\Sigma$ is the 
exchange cross-section. 
Consider the likelihood of a BS of mass $M_3 = 2.0 \Msun$ being exchanged 
into a binary with $M_1 = 1.0 \Msun$ and $M_2 = 0.5 \Msun$. 
From the results of binary-single-star scattering experiments performed by 
Heggie, Hut \& McMillan~(1996) the exchange cross-section to displace 
$M_1$ is $13\,300 \, a \, {\rm AU}^2$ and to replace $M_2$ is 
$35\, 400 \, a \, {\rm AU}^2$.  
Taking $n_{\rm b} = 20 \, {\rm pc}^{-3}$, $V_{\rm rel} = 1 \, {\rm km} \, 
{\rm s}^{-1}$ and $a = 100\,$AU gives ${\tau}_{\rm ex} \simeq 540\,$Myr which 
is an order of magnitude less than the age of M67. 
On the other hand, the tidal capture timescale for the BS is about $10^4\,$Myr 
(Press \& Teukolsky 1977). 
This is confirmed by Portegies Zwart et al.~(1997) who find that in a cluster 
core with $\log \left( n/{\rm pc}^{-3} \right) = 3.92$ the low encounter 
rate means that tidal capture is rare. 
We must however be careful with the definition of tidal capture because some 
BS binaries in the M67 model do form from bound triple systems which 
themselves formed from a binary-single-star interaction. 

Even though during the M67 simulation two BSs, formed from a collision within 
a triple system, are found in short-period eccentric orbits it is 
hard to see how the cluster dynamics can produce significant numbers of 
these binaries. 
The exchange cross-section is proportional to the binary separation 
(Heggie, Hut \& McMillan 1996) so they are unlikely to be formed in this 
way (see Figure~\ref{f:fig8bs}). 
Allowing for the chance formation of a tidal capture binary,  
the calculations of Portegies Zwart et al.~(1997) show that capture 
binaries in a dense cluster core are generally close with $a < 10 \Rsun$ 
but that 60\% of these circularize during the formation process. 
So for every BS observed in a close eccentric orbit at least one should 
be found in a close circular orbit. 
BSs in close circular orbits can be formed from Case~B mass transfer but 
these are even less likely to have an eccentricity induced than wide 
circular binaries. 
A possibility that has yet to be considered and does not rely on cluster 
dynamics is formation via 
common-envelope evolution when unstable Case~B mass transfer occurs. 
The processes involved in common-envelope evolution are very uncertain 
(Iben \& Livio 1993) and there is no real reason to assume that 
a binary emerging from this phase should be circular. 
Therefore the binary S1284 observed in M67 could be produced in this way 
or possibly by a collision within a triple system. 

The number of blue stragglers produced by the M67 model is in good agreement 
with the observations. 
Figure~\ref{f:fig1bs} shows that the cluster environment is very effective 
at increasing the relative number of the BS population.  
If it had been possible to start a full simulation from $T = 0$ then 
hopefully the relative number of BSs would 
increase gradually with time to reach the M67 point, rather than the 
rapid increase produced by the semi-direct simulation (although the 
log-scale in Figure~\ref{f:fig1bs} distorts the actual range of the 
simulation).  
As expected the BSs formed by a variety of paths, Case~A and Case~B mass 
transfer in binaries with orbital parameters unaffected by the cluster 
environment, Case~A mass transfer after perturbation-induced 
circularization, and collisions in highly eccentric binaries. 

In population synthesis without cluster dynamics the dominant formation 
mechanism is Case~A mass transfer. 
However, Mathys~(1991) has observed that M67 BSs rotate slower than 
average for MS stars. 
As noted by Leonard \& Linnell~(1992), tidal synchronization in a close 
binary should cause BSs produced by Case~A mass transfer to be rapid 
rotators. 
Magnetic braking will not be effective at reducing the rotation rate because  
MS stars with $M \simgtr 1.3 \Msun$ do not have convective envelopes. 
It is uncertain whether BSs resulting from collisions between two MS stars 
will be rapid rotators but it is interesting that 50\% of the BSs in the 
$N$-body model came from direct collisions in eccentric binaries. 
The progenitor stars will not have been affected by standard tidal 
circularization in this case but it is unclear what the effect of any 
angular momentum exchange during chaotic motions of the orbit will 
have on the stellar spins.  
Our simulation did not form any BSs from hyperbolic collisions between two 
single stars, in agreement with the collision timescale predicted by Press \& 
Teukolsky~(1977), which is about $10^6\,$Myr for a $1 \Msun$ star in the core 
of M67.  

An encouraging number of wide eccentric BS binaries are 
formed during our simulation. 
Super-BSs are also made, as well as some short period binary BSs in eccentric 
and circular orbits. 
So the model contains formation paths for all the BSs observed in M67,  
except wide circular BS binaries but this can be rectified as discussed above.  
Another discrepancy is that the frequency of model BSs found in binaries is too 
low.  
The BS half-mass radius in the model is smaller than the half-mass radius of 
the MS stars, in qualitative agreement with the observations, but is itself 
a factor of 2 less than is observed for M67. 
Also too many potential RS$\,$CVn systems are broken-up in the cluster core 
which suggests that the population within the half-mass radius is too 
centrally condensed. 
Observations of M67 suggest that the number density of stars is about 
$n = 40 \, {\rm pc}^{-3}$ within a core of radius $1\,$pc, while the model at 
$4\,300\,$Myr has $n = 150 \, {\rm pc}^{-3}$ within $1\,$pc of the cluster 
centre and $n = 34 \, {\rm pc}^{-3}$ inside the half-mass radius. 
A decrease in the central concentration implies that wide binaries 
formed from exchange interactions have longer lifetimes. 

The use of a rather unusual tidal field in the M67 model, 
prompted by the observed boundary of $10\,$pc for the cluster stars,  
deserves further discussion. 
Preferential escape of low-mass stars from the model owing to two-body 
encounters shows that the current observed mass of M67, for stars with 
masses greater than $0.5 \Msun$, may be close to the actual cluster mass,  
and therefore a standard tidal field may well apply. 
The case for this is strengthened when we consider that the observed 
boundary of the cluster members is only a lower limit for 
the tidal radius. 
However, it is interesting that 
$v_{\rm\SSS G} = 350 \, {\rm km} \, {\rm s}^{-1}$ is not ruled out by 
the model of the Milky Way halo presented by Wilkinson \& Evans (1999). 
Additionally, 
Baumgardt (1998) has shown that it is the tidal radius at perigalacticon 
of an eccentric orbit that determines the dissolution of a star cluster. 
The main effect of a stronger tidal field is to drive the cluster 
evolution at a higher rate. 
Therefore it is possible that use of a standard tidal field would 
cause the peak in $N_{\rm\SSS BS}$ that we see in Figure~\ref{f:fig7bs} 
to occur at a later time, closer to the age of M67. 
It is also possible that as the core would not be as dynamically evolved 
it would be less dense, leading to a greater population of wide binaries 
and RS$\,$CVn systems. 
We should stress that these points are purely conjecture and that  
the tidal field requires close attention in future simulations. 
In particular, work is currently underway to implement a time-varying 
tidal field in {\tt NBODY4} which will enable eccentric cluster orbits 
to be followed (Wilkinson \& Hurley 2001). 

Ideally it would be desirable to perform more simulations and 
investigate the effects of varying parameters such as the central 
concentration in the starting model.  
Many factors are uncertain, both in the model and the observations. 
For example, 
the derived escape rate may not be correct for larger $N$ so the size of 
the starting model, in terms of both star number and length scale, may 
not be relevant to the conditions of M67 at birth. 
In all of this it would be helpful if we could begin a complete simulation 
from initial conditions but, considering that a single semi-direct simulation  
took a month to perform, the capability of the current hardware is already 
pushed to its limit. 
Future improvements in computing efficiency, such as the availability of
GRAPE-6 (Makino 1999), will enable a leap forward in the field of
star cluster modelling.
The ability to perform direct simulations of clusters like M67 comfortably 
will decrease substantially the uncertainties involved,
and allow observed and model parameters to be matched iteratively. 

Finally we also need an explanation for the two subsubgiant binaries 
observed in M67, S1063 and S1113. 
These lie below the base of the GB (BGB) in the CMD and to the right of the 
MS, much further displaced than the binary MS. 
S1063 has orbital parameters $P = 18.4\,$d and $e = 0.217$ while S1113 
is circular with $P = 2.8\,$d. 
HG stars respond to changes in mass on a thermal timescale 
(see Section~2.7 of Hurley, Pols \& Tout~2000) so that as they lose mass 
they will evolve below the HG of the cluster isochrone. 
Mass transfer is consistent with the orbit of S1113 but not the eccentric orbit 
of S1063, unless the observed eccentricity could be for the outer 
orbit of a triple system. 
The population synthesis run PS6 produces 30 binaries in the subsubgiant 
area at $4.2\,$Gyr per $500\,000$ evolved. 
These are all circular, have a HG primary of average mass $1 \Msun$, and an 
average period of $2\,$d. 
So if M67 originally had $15\,000$ binaries then one binary with these orbital 
parameters can be expected, i.e. S1113. 

Possible evolution paths for S1063 are harder to find. 
The sub-luminous star may have formed in the same way as for S1113 and then 
been exchanged into an eccentric binary. 
However it is unlikely that two HG primaries are transferring mass at the same 
time and the exchange timescale for a short-period binary is long as has 
already been discussed. 
Another explanation is that a MS star with $M < M_{\rm TO}$ is evolving 
off the MS before it should.  
Consider a primordial binary with $M_1 = 1.4 \Msun$, $M_2 = 0.4 \Msun$, 
$P = 10\,$d and $e = 0.9$. 
At $T = 2\,500\,$Myr the more massive star begins to transfer mass to its 
companion. 
The normal MS lifetime of the primary is $3\,370\,$Myr but due to the mass 
transfer it ages as it loses mass and at $2\,700\,$Myr, 
when $M_1 = 1.2 \Msun$, its effective MS lifetime is $4\,200\,$Myr. 
If for some reason the mass transfer were halted at this point then the 
star would evolve off the MS at $4\,200\,$Myr into the subsubgiant region. 
Exchange of the star into a wider binary is one way to halt the mass transfer 
and to explain the observed parameters of S1063. 
In run PS6 there are 600 likely exchange targets per $500\,000$ binaries 
but the window for exchange is small, as it must occur when $T_{\rm MS} 
\simeq 4\,200\,$Myr, and so is the exchange cross-section. 
The exact nature of S1063 remains a mystery. 

\section{Conclusions}
\label{s:m67con}

In order to complement increases in computing speed we have made a 
substantial effort to improve the treatment of physical processes 
within the cluster environment. 
Our $N$-body code includes detailed modelling of stellar and binary evolution,  
thus allowing us to test directly the influence of the interaction 
between stellar evolution and gravitational encounters on the evolution
of a star cluster.
Some similar advances have also been made in the work of
Portegies Zwart et al~(1998, 2000) although their models still only include 
Population I evolution and are therefore not suited to studying 
globular clusters.
In particular, certain aspects of binary evolution, such as tidal 
circularization, are not modelled in as much detail as this work. 

{\tt NBODY4} is extremely useful for simulating cluster populations. 
As a first application we have modelled the blue straggler population 
in M67. 
We could just as easily have looked at many other aspects of cluster 
evolution, such as the production of CVs or development of mass segregation, 
but such topics will be the subject of future work. 
In particular, the availability of GRAPE-6 will enable us to 
model small globular clusters directly. 

We have shown that binary evolution alone cannot account for the numbers 
of observed blue stragglers in open clusters, or the binary properties 
of these blue stragglers, when a realistic separation distribution is assumed. 
The influence of the cluster environment can effectively 
double the number of blue stragglers produced, leading to good agreement with 
the observations. 
Our $N$-body model of M67 demonstrates that blue stragglers are most likely 
generated by a variety of processes and in particular 
we find formation paths for all the BSs observed in M67. 
We also find that, among the possibilities we consider, 
the primordial binary population is best 
represented by a log-normal distribution of separations peaked at $10\,$AU 
and binary masses chosen from the mass function of Kroupa, Tout \& Gilmore 
(1991) in combination with a uniform mass-ratio distribution. 

We quantify the escape rate of stars from a cluster subject to the tidal field 
of our Galaxy as $\dot{M} = - 0.3 M / (t_{\rm rh} \, {\log}_{10} 0.4N)$. 
This enables us to provide a method by which the initial cluster mass and 
radius corresponding to current values can be determined. 
So we were able to model M67 by a semi-direct method 
even with current computational limitations.

\section*{ACKNOWLEDGMENTS}

JRH thanks Trinity College and the Cambridge Commonwealth Trust for their
generous support.
CAT is very grateful to PPARC for support from an Advanced Fellowship.
ORP thanks the Institute of Astronomy, Cambridge for supporting a number of
visits undertaken during this work.
We thank Douglas Heggie, Rainer Spurzem and Philipp Podsiadlowski 
for many helpful comments that improved the quality of this manuscript. 
We also thank the referee (Mirek Giersz) for his suggestions on a 
number of important points.

\label{lastpage}
\end{document}